\documentclass[prc,aps,amssymb,amsmath,amsfonts,11pt]{revtex4-1}
\usepackage{latexsym}
\usepackage[latin1]{inputenc}
\usepackage[dvips]{graphicx}
\usepackage{color}
\usepackage{changes}
\definechangesauthor[name={Juan}, color=blue]{JT}

\newcommand{\ba}{\begin{eqnarray}}
\newcommand{\ea}{\end{eqnarray}}

\newcommand{\be}{\begin{equation}}
\newcommand{\ee}{\end{equation}}
\newcommand{\pa}{\partial}

\usepackage{dcolumn}

\begin{document}
                    
\title{Electrical conductivity and relaxation via colored noise in a hadronic gas}

\author{Jan Hammelmann$^{1,2}$, Juan M. Torres-Rincon$^{3}$, Jean-Bernard Rose$^{1,2}$, Moritz Greif$^{2}$, Hannah Elfner$^{4,2,1}$}

\affiliation{$^1$Frankfurt Institute for Advanced Studies, Ruth-Moufang-Strasse 1, 60438
Frankfurt am Main, Germany}
\affiliation{$^2$Institute for Theoretical Physics, Goethe University,
Max-von-Laue-Strasse 1, 60438 Frankfurt am Main, Germany}
\affiliation{$^3$Department of Physics and Astronomy, Stony Brook University,
Stony Brook, New York 11794-3800, USA} 
\affiliation{$^4$GSI Helmholtzzentrum f\"ur Schwerionenforschung, Planckstrasse 1, 64291
Darmstadt, Germany}

\begin{abstract}
Motivated by the theory of relativistic hydrodynamic fluctuations we make use of the Green-Kubo formula to compute the electrical conductivity and the (second-order) relaxation time
of the electric current of an interacting hadron gas. We use the recently developed transport code SMASH to numerically solve the coupled set of Boltzmann
equations implementing realistic hadronic interactions. In particular, we explore the role of the resonance lifetimes in the determination of the electrical relaxation time. As opposed to a previous 
calculation of the shear viscosity we observe that the presence of resonances with lifetimes of the order of the mean-free time does not appreciably affect the relaxation of the electric current fluctuations.
We compare our results to other approaches describing similar systems, and provide the value of the electrical conductivity and the relaxation time for a hadron gas at temperatures between $T=60$ MeV and $T=150$ MeV.
\end{abstract}

\maketitle

\section{Introduction}

In high-energy collisions of heavy nuclei, such as those at the Large Hadron Collider (LHC) or the Relativistic Heavy-Ion Collider (RHIC) facilities, a transient state of hot deconfined matter is created. It consists of a strongly interacting quark-gluon plasma (QGP), which expands and cools down in time, until a transition temperature ($T_c$) is reached and the
medium hadronizes~\cite{Arsene:2004fa,Adcox:2004mh,Back:2004je,Adams:2005dq}. The hadron gas is still hot, and interacts by collisions and the exchange of resonances. At the same time it keeps expanding and cools further until complete chemical and kinetic freeze out. Much later, detectors register hadronic spectra and analyze different observables. Low energy heavy-ion collisions (HICs), such as those at Super Proton Synchrotron, the RHIC Beam Energy Scan~\cite{Mitchell:2012mx,Mohanty:2011nm}, or at GSI~\cite{Ablyazimov:2017guv} or 
NICA~\cite{Kekelidze:2012zz}, the QGP phase is very short---or not even reached~\cite{Heinz:2000bk}---, but the hadron gas dynamics is still present and dominates the evolution of the system. As this low-energy phase of quantum chromodynamics (QCD) is nonperturbative, practitioners usually make use of effective models like hydrodynamics~\cite{Shen:2012vn,Kolb:2003dz,Teaney:2001av}, transport theory~\cite{Weil:2016zrk,Buss:2011mx,UrQMD1,Xu:2004mz,Cassing:2009vt} or fireball~\cite{Rapp:2014hha} descriptions.

In order to characterize the interacting hadron gas in a simple way, meaningful quantities are linear response transport coefficients, such as shear and bulk viscosities, diffusion coefficients or conductivities. Those quantities map out the interactions strength of medium constituents, as well as their (relative) abundances and charges. It is
desirable to understand their values and dependencies, both for the characterization of matter and to provide robust inputs for hydrodynamical models.

Numerical solutions of transport equations allow us to extract these transport coefficients, but  require in turn microscopic interaction probabilities as input, which are in many cases not known precisely from experiments, or again rely on models. For instance, hadronic cascades were extremely successful in the past in explaining a variety of experimental observables both for high and low energy collisions~\cite{UrQMD1,UrQMD2,GiBUU,Weil:2016zrk}. This, in turn, is a useful tool to learn about interactions among hadrons and resonances. Transport coefficients from numerical transport models, benchmarked by experiments, supplements insufficient results from lattice QCD or Dyson-Schwinger equations and is a practical way to infer QCD properties in the nonperturbative, confined regime. The use of transport equations, when constructed from the corresponding low-energy field theory, would provide consistent information about nonequilibrium properties. In our approach, we will follow a semiclassical perspective, based on the Boltzmann transport equation.

In this work, we employ the hadronic cascade ``Simulating Many Accelerated Strongly-interacting Hadrons'' (SMASH)~\cite{Weil:2016zrk} to continue our study of transport coefficients in an equilibrated 
hadronic gas using the Green-Kubo formalism. Thus we follow the path started in our previous works,~Refs.~\cite{Rose:2017ntg,Rose:2017bjz,Wesp:2011yy} focusing on shear viscosity, but now compute the electrical conductivity.
Since SMASH constitutes a good description of various experimental data, like hadron properties and cross sections~(see \cite{Weil:2016zrk,Staudenmaier:2017vtq} for more details, and~\cite{github,analysisresults} for on-line examples of its performance), the value of the transport coefficients extracted in this work can be seen
as a result restricted by experiment. 

After presenting essential numerical tests, where we will compare to kinetic theory (mostly analytic) calculations~\cite{Greif:2016skc}, we investigate in detail the role of resonances. As observed in our previous study, Refs.~\cite{Rose:2017ntg,Rose:2017bjz}, the lifetimes of the formed resonances might influence the relaxation of the fluctuations, affecting the transport coefficient at high temperatures.
This is an important dynamical feature of the propagating resonances, which is absent in all previous studies based on semianalytical solutions of transport equations. Our final goal is to present a 
solid determination of the electrical conductivity of a realistic hadronic gas, taking into account these effects.  

The electrical conductivity describes the response of a medium to an external electric field or an accumulation of charge density. It is only sensitive to transport cross sections involving charged particles, as opposed to, e.g., shear viscosity which obtains equal 
contributions of charged and uncharged particles. It is furthermore interesting to remark that the low-mass dilepton yield is proportional to the electric
conductivity, which can provide additional experimental cross checks~\cite{Moore:2006qn,Ding:2016hua,Ghiglieri:2016tvj}. It can also be related to the diffusion of 
magnetic fields in a medium~\cite{Baym:1997gq,Tuchin:2013ie,FernandezFraile:2005ka} and is an important input to magnetohydrodynamics~\cite{Roy:2017yvg,Inghirami:2016iru,Roy:2015kma},
allowing for a longer duration of the initial magnetic field when this coefficient is finite (as opposed to the ideal magnetohydrodynamic case).

The electrical conductivity has gained increasing interest over the past years, since it became more and more apparent that electric and magnetic fields in high energy collisions are very strong in the early phase~\cite{Greif:2017irh,Huang:2015oca}. Their dynamics depends by Maxwell's equations  on the
electrical conductivity. This coefficient was computed in hadronic kinetic theory~\cite{Greif:2016skc},
partonic transport models~\cite{Greif:2014oia,Puglisi:2014sha}, off-shell transport and dynamical quasiparticle
models~\cite{Cassing:2013iz,Steinert:2013fza,Marty:2013ita,Berrehrah:2015vhe}, holography \cite{Finazzo:2013efa,Finazzo:2014cna,Rougemont:2017tlu}, lattice QCD~\cite{Amato:2013naa,Aarts:2014JHEP,Brandt:2015aqk,Ding:2016hua}, Dyson-Schwinger
calculations \cite{Qin:2013aaa}, in the Polyakov-extended quark-meson model~\cite{Singha:2017jmq}, semianalytic calculations within perturbative QCD~\cite{Arnold:2000dr,Arnold:2003zc,Mitra:2016zdw}, and also taking into account strong magnetic fields~\cite{Hattori:2016cnt,Fukushima:2017lvb}. Closer to the systems we are considering in this work we mention semianalytical 
calculations in a pion gas using chiral perturbation theory with and without unitarization~\cite{FernandezFraile:2005ka, Torres-Rincon:2012sda}, in a pion gas using a sigma model with and without medium-modified 
interactions~\cite{Ghosh:2018kst}, in a sigma model with baryon interactions~\cite{Ghosh:2016yvt}, and in resonance gas models~\cite{Kadam:2017iaz,Greif:2016skc}.

This work is organized as follows. After introducing the theory of relativistic colored hydrodynamic fluctuations and the Green-Kubo formalism in Sec.~\ref{sec:1}, we present the numerical framework SMASH in Sec.~\ref{sec:2}, and general considerations about methodology in Sec.~\ref{sec:3}.
In Sec.~\ref{sec:4} we investigate simpler test-cases and verify our numerical method, followed by our results for $2\leftrightarrow 2$ scattering in Sec.~\ref{sec:5}. In Sec.~\ref{sec:6} we present the results for
the full interacting hadron gas and conclude in Sec.~\ref{sec:7}.

\section{Electric current fluctuations and Green-Kubo formula~\label{sec:1}}

Relativistic hydrodynamics is an effective description for the long-wavelength modes of a relativistic fluid. It assumes local equilibration (although its validity might be more general~\cite{Florkowski:2017olj,Romatschke:2017ejr}), 
and it contains the equations of motion of a small set of hydrodynamic fields, like the fluid's velocity, particle and energy densities.
As a result, the otherwise complicated equations of motion of particles are traded by a small number of conservation laws, providing a huge simplification of the many-body problem.

As global equilibrium is not required in hydrodynamics, small gradients might exist in the system. The nonequilibrium currents (responses of the fluid) are expanded in powers of gradients
of the hydrodynamic fields, and the respective coefficients are the transport coefficients, which have microscopical origin. For instance, the energy-momentum tensor admits the expansion
(using the Landau frame)
\be T^{\mu \nu} (t,{\bf x})= T_{\textrm{ideal}} + \eta \left( \nabla^\mu u^\nu + \nabla^\nu u^\mu \right) + \left( \frac{2}{3} \eta -\zeta \right) \Delta^{\mu \nu} \partial_\alpha u^\alpha \ ,  \ee
where the ideal term is $T_{\textrm{ideal}}^{\mu \nu}=diag(\epsilon,P,P,P)$ in the local rest frame and
$\Delta^{\mu \nu}=u^\mu u^\nu -\eta^{\mu \nu}$ and $\nabla^\mu=-\Delta^{\mu \nu} \partial_\nu$ are the spatial projector and derivative, respectively. 
We use the Minkowski metric $\eta^{\mu \nu}$ with the mostly-minus convention. 

  In this paper we focus on systems containing electrically charged particles. The conservation of the $U(1)$ electric charge $Q$ introduces an additional transport coefficient which 
represents the ability of the system to create an electric current in the presence of an external electric field~\footnote{Other conserved charges, like baryon and strangeness charge, can induce cross-diffusion and mixing effects~\cite{Greif:2017byw} which are out of scope of this paper but can be investigated with similar methods.}. This coefficient is called the electrical conductivity $\sigma_{\textrm{el}}$ (we consider an
isotropic medium, where this coefficient is a scalar quantity). The corresponding phenomenological relation is the relativistic version of the Ohm's law~\cite{landau1981course}
\be J_Q^\mu (t,{\bf x}) = \sigma_{\textrm{el}} E'^\mu (t,{\bf x}) \ , \label{eq:ohm} \ee
where $J_Q^\mu (t,{\bf x})$ denotes the 4-current related to the $U(1)$ gauge symmetry (electromagnetism), $E'^\mu \equiv E^\mu -T \nabla^\mu (\mu_Q/T)$, with $E^\mu=F^{\mu \nu} u_\nu$ the electric field,
$F^{\mu\nu}$ the electromagnetic field-strength tensor, $T$ the temperature, and $\mu_Q$ the charge chemical potential.

In the local rest frame, $u^\mu=(1,0,0,0)$, Ohm's law reduces to the well-known version ${\bf J}_Q (t,{\bf x})=\sigma_{\textrm{el}} {\bf E'} (t,{\bf x})$. The electrical conductivity $\sigma_{\textrm{el}}$ can be
calculated via the Green-Kubo formalism using the current-current correlation function~\cite{kubo}, which can be obtained from linear response theory~\cite{zubarev,landau1981course}. However, as we will 
study the response of the system to internal fluctuations (any external perturbation will be absent) we will provide a different---but entirely equivalent---derivation using the theory of
hydrodynamic fluctuations~\cite{landaufluc, Kapusta:2011gt}.

  Equation~(\ref{eq:ohm}) is only true on average, because microscopic motion induces small fluctuations around Ohm's law~\cite{landaufluc, boon1980molecular, Kapusta:2011gt},
\be \label{eq:jQ} J_Q^i (t,{\bf x}) = \sigma_{\textrm{el}} E'^i (t,{\bf x}) + \xi_Q^i (t,{\bf x}) \ , \ee
where $\xi_Q (t,{\bf x})$ is the hydrodynamic noise of the electric current due to fast microscopic dynamics. In particular, Eq.~(\ref{eq:jQ}) states that a nonzero electric current will be produced even in the absence of
an external electric field. Only after repeated measurements, the electric current will approach, on average, the Ohm's value. Then, the thermal noise satisfies
\be \langle \xi_Q^i (t, {\bf x}) \rangle = 0 \ , \ee
where the brackets indicate an average over the ensemble of experimental measurements. However, the 2-point correlation function is not zero due to the fluctuation-dissipation theorem~\cite{Kapusta:2011gt},
\be \label{eq:flucdis} \langle \xi_Q^i(t, {\bf x}) \xi_Q^j(t',{\bf x'}) \rangle = 2T \sigma_{\textrm{el}} \delta^{ij} \delta(t-t') \delta^{(3)} ({\bf x}-{\bf x'}) \ . \ee 
Therefore, this approach assumes a white Gaussian noise (we will shortly see that this is too restrictive in connection with the microscopic description).

  In global equilibrium over a volume $V$, without any external gradients, we can define the average electric current ${\bf j}_Q(t)=\frac{1}{V} \int d^3x {\bf J}_Q (t,{\bf x})$, whose correlation function reads
\be \langle j_Q^i(t) j_Q^j(t') \rangle = \frac{2T\sigma_{\textrm{el}}}{V} \delta^{ij} \delta(t-t')  \ . \ee
We notice that the correlation of the electric current is local in time, and that the strength of the fluctuation $\sqrt{ j^i_Q(t) j^i_Q(t)}$ is proportional to $\sqrt{T/V}$.

Setting $t>t'=0$ and integrating over time we observe that this form of the noise is compatible with the Green-Kubo formula for the electrical conductivity~\cite{kubo,zubarev},
\be \label{eq:GK} \sigma_{\textrm{el}} = \frac{V}{3T} \int_0^{\infty} dt \ \langle {\bf j}_Q (t) \cdot {\bf j}_Q (0) \rangle \ . \ee
  
  \subsection{Colored noise}
  
  Equation~\eqref{eq:ohm} is known to be inconsistent with causality as a sudden application of an external electric field would create an instantaneous electric current. 
Retardation effects are missing in the constitutive law. According to this, the nonequilibrium electric current should relax toward Ohm's law after some time. Ohm's law is modified to
\be J_Q^i (t,{\bf x}) =\int_V d^3\mathbf{x^\prime} \int_{t_0}^{t} dt' \ \Sigma^{ij} (t-t',{\bf x}-{\bf x}') E'^j (t',{\bf x}') + \Xi_Q^i (t,{\bf x}) \ , \label{eq:ohmcausal} \ee
where $\Sigma^{ij} (t-t',{\bf x}-{\bf x}')$ is the memory kernel for the electrical conductivity.

For dilute systems, the so-called relaxation time approximation (RTA) of the Boltzmann equation assumes an exponential decay of the initial perturbation of the distribution function, and therefore, the same decay of the nonequilibrium current. This approximation motivates the use of an exponential {\it ansatz} for the memory kernel (we will check this assumption {\it a posteriori}),
\be \label{eq:expo} \Sigma^{ij} (t-t',{\bf x}-{\bf x}')=\frac{\sigma_{\textrm{el}} \delta^{ij}}{\tau_Q} \delta^{(3)}({\bf x}-{\bf x}') \exp \left( - \frac{|t-t'|}{\tau_Q} \right) \ , \ee
where $\tau_Q$ is a relaxation time. 

The prefactors of this kernel are chosen in such a way that when $\tau_Q \rightarrow 0$ one obtains,
\be \lim_{\tau_Q \rightarrow 0} \Sigma^{ij} (t-t',{\bf x}-{\bf x}') = 2 \sigma_{\textrm{el}} \delta^{ij} \delta^{(3)} ({\bf x} - {\bf x'}) \delta(t-t') \ , \ee
and one recovers the standard Ohm's law in that limit.

Equation~(\ref{eq:ohmcausal}) is compatible with causality, but applying the fluctuation-dissipation theorem it renders the noise nonwhite~\cite{Murase:2013tma},
\be \label{eq:colored} \langle \Xi_Q^i(t,{\bf x}) \ \Xi_Q^j (t',{\bf x}') \rangle = T \Sigma^{*,ij} (t-t',{\bf x}-{\bf x}') = \frac{\sigma_{\textrm{el}} T}{\tau_Q} \delta^{ij} \delta^{(3)} ({\bf x} - {\bf x'}) \exp \left( - \frac{|t-t'|}{\tau_Q} \right) \ . \ee

The use of colored noise will be an essential property for the connection with the microscopic description.
The presence of colored noise at small timescales comes from the simple fact that the electric current is determined from a coarse-graining procedure of the microscopical description, and the particle properties 
are correlated in time of the order of the mean-free path ($\lambda_{\mathrm{mfp}} \lesssim \tau_Q$).

It is very instructive to realize that Eq.~(\ref{eq:ohmcausal}) can be shown to be equivalent to the following law~\cite{Murase:2013tma,Kapusta:2017hfi}
\be \label{eq:colored2} \tau_Q \frac{\pa J^i_Q(t,{\bf x})}{\pa t} + J^i (t,{\bf x}) =  \sigma_{\textrm{el}} E'^i (t, {\bf x})  + \xi_Q^i (t,{\bf x}) \ ,  \ee
where the fluctuating term $\xi_Q^i (t,{\bf x})$ is now a white noise satisfying Eq.~(\ref{eq:flucdis}). This brings two important conclusions:
\begin{enumerate}
 \item Colored Gaussian noise makes the phenomenological Ohm's law consistent with second-order causal hydrodynamics (whenever the relaxation time is large enough). 
 \item The coefficient $\tau_Q$ appearing in the noise correlation function has the interpretation of the electric current relaxation time. It appears in causal extensions of hydrodynamics like 
 the Maxwell-Cattaneo equation. See also the Appendix of~\cite{Kapusta:2012zb} for a Lorentz invariant version of this expression for the baryon current and~\cite{Kapusta:2017hfi} for an application of this colored noise in HICs.
\end{enumerate}

 In Appendix~\ref{app} we illustrate the restoration of causality due to $\tau_Q$ for a particular 1+1 dimensional evolution of an initial local charge fluctuation with the transport coefficients found using SMASH. 
 
 Finally, we make note that the exponential ansatz is compatible with the Green-Kubo approach. In the absence of external fields, the correlation function of the spatial-average electric current is, from Eq.~\eqref{eq:colored},
\be \langle j_Q^i (t) j_Q^j (t') \rangle=  \frac{\sigma_{\textrm{el}} T \delta^{ij}}{\tau_QV} \exp \left(- \frac{|t-t'|}{\tau_Q} \right)   \label{eq:corrcausal} \ . \ee
After setting $t'=0$ and integrating over $t$ we recover the Green-Kubo relation in Eq.~(\ref{eq:GK})
\be \sigma_{\textrm{el}} = \frac{V}{3T} \int_0^{\infty} dt\, \langle {\bf j}_Q (t) \cdot {\bf j}_Q (0) \rangle =\frac{V}{T} \int_0^\infty dt\, \langle j_Q^x(t) j_Q^x(0) \rangle \ , \ee
assuming isotropy of the system in the last step.

Exploiting the exponential {\it ansatz} we obtain,
\be \label{eq:expansatz} C(t) \equiv \langle j_Q^x (t) j_Q^x (0) \rangle= C(0) \exp \left( - \frac{t}{\tau_Q} \right) \ .  \ee
so that the electrical conductivity can be readily computed as
\be \label{eq:conducsimp} \sigma_{\textrm{el}} = \frac{\tau_Q C(0)V}{T} \ . \ee
It is worth mentioning that the formula (\ref{eq:conducsimp})
with the value of $C(0)V$ given in~(\ref{eq:preC0}) is very similar to the expression 
given in the context of the Kubo relation in~\cite{FernandezFraile:2005ka}.
Apart from the different statistics (Boltzmann versus Bose-Einstein), the main difference is that the role of our $\tau_Q$ is played by the inverse pion width $\Gamma_p^{-1}$. In the context of the chiral perturbation theory, this pion width can be computed as the imaginary part of the pion self-energy (at lowest order, the 2-loop ``sunset'' diagram). It is well known~\cite{Weldon:1983jn} that at finite temperature this imaginary part of the self-energy has the interpretation of the relaxation rate of the distribution function towards equilibrium.
Using such distribution function to describe the non-equilibrium electric current (by performing the integration over the phase space with the appropriate weight), the inverse pion width appears as the relaxation time of $J^i_Q$, thus matching our interpretation of $\tau_R$ in (\ref{eq:colored2})
(notice that this identification works only in the dilute regime; in general, the two times $\tau_Q$ and $\Gamma_p^{-1}$ are not the same as they have very different interpretations).

In this work we use the relativistic microscopical hadron transport model SMASH to compute $\tau_Q$ and $C(0)$ for a given system~\footnote{$C(0)$ is easily analytically computable but can be cross-checked by SMASH.} with fixed $V$ and $T$. In particular, we will show that hydrodynamic fluctuations---directly computed after the coarse graining procedure---follow a colored distribution, whose correlation function follows the exponential {\it ansatz} in a very good approximation.

We compute the electrical conductivity and the relaxation time of the electric current for several simple systems at finite temperature and chemical potential, while providing several consistency tests of our approach. We will describe the differences between the transport calculation in SMASH based on resonance propagation and analytical solutions of the Boltzmann equation obtained for similar systems.
Such differences have been recently addressed for the shear viscosity in Ref.~\cite{Rose:2017bjz} and ascribed to the interplay of the resonance lifetime and mean-free time. We study the same effects here, and finally 
present the temperature and chemical potential dependence of the electrical conductivity for the hadronic medium.
  
\section{SMASH~\label{sec:2}}
  
In this work we address the hydrodynamic fluctuations and the computation of the electrical conductivity by addressing the dynamics of individual particles simulated in a transport model. We use the recently developed transport model SMASH~\cite{Weil:2016zrk}. 
It is applicable for the full evolution of the system created in heavy-ion collisions at SIS-18 energies~\cite{Staudenmaier:2017vtq,Steinberg:2018jvv}, while at high RHIC and LHC energies it has been applied for the late stage rescattering dynamics~\cite{Oliinychenko:2018ugs}.

 It effectively solves the set of coupled Boltzmann equations~\cite{de1980relativistic} for the different hadron species,
\begin{equation} \label{eq:boltzmann}
 p^\mu \frac{ \partial f_i (t,{\bf x},{\bf p})}{\partial x^\mu} + m_i F^{\mu} \frac{ \partial f_i (t,{\bf x},{\bf p})}{\partial p^\mu} = C_{\textrm{coll}} [f_i] \ ,
\end{equation}
where $f_i$ is the one-particle distribution function of the species $i$, $F^\mu=F^\mu ({\bf x},{\bf p})$ is an external force (set to zero in this work), and $C_{coll}$ is the nonlinear collision integral. It describes the interactions of the $i$-th particle with the rest. The distribution function is evolving in time corresponding to this equation of motion. 

The interaction with other particles is constructed via a geometric collision criterion
\begin{equation} \label{eq:geometriccriterion}
d_{\textrm{coll}} < \sqrt{\frac{\sigma_{\textrm{tot}}}{\pi}} \ ,
\end{equation}
where $d_{\textrm{coll}}$ is the relative distance between the two particles and $\sigma_{tot}$ the total cross section.
The particles existing in SMASH are well-known hadrons taken from Ref.~\cite{Patrignani:2016xqp}, with their respective mass and decay width. For collision energies with $\sqrt{s_{NN}}$ of a few GeV the
interaction between hadrons can be modeled by formation of resonances. The energy dependent cross section follows a Breit-Wigner distribution function with a mass-dependent width and peaks at the pole mass of the resonance. Unstable particles will decay after propagating a certain lifetime into allowed decay channels.
In this work all particles we used are summarized in Table~\ref{tab} with their corresponding masses and decay widths. Isospin symmetry is assumed.

\begin{table}[ht]
\begin{center}
\begin{tabular}{|c|c|c|}
  \hline
    Particle & Mass (MeV) & Decay width (MeV) \\
  \hline
  \hline
    $\pi$ & $138$ & $0$ \\
  \hline
    $\rho$ & $776$ & $149$ \\
  \hline
    $K$ & $494$ & $0$ \\
  \hline
    $K^\star$ & $892$ & $50.8$ \\
  \hline
    $N$ & $938$ & $0$ \\
  \hline
    $\Delta$ & $1232$ & $117$ \\
  \hline
\end{tabular}
\caption{Masses and decay widths of the hadrons used in this work as components of the hadronic gas.\label{tab}}
\end{center}
\end{table}

\section{Methodology~\label{sec:3}}
\label{sec:2.5}
As explained in Sec.~\ref{sec:1} we employ the Green-Kubo formalism in order to compute the electrical conductivity. This approach is well known and has been used in many calculations before~\cite{Demir:2008tr,Wesp:2011yy,Greif:2014oia,Puglisi:2014sha,Rose:2017bjz}.
For all systems the electrical conductivity is calculated in a cubic box of fixed volume where we measure the properties of particles. We apply periodic boundary conditions to avoid wall effects and no external electric field is applied.
The determination of the equilibrium properties (temperature and baryon density) is common to all our examples and it follows the same procedure used in our previous study of the shear viscosity~\cite{Rose:2017bjz}. 
Due to the necessity of the thermal equilibrium of the system, both thermal and chemical equilibria of the boxes have been tested, before calculating the electrical conductivity $\sigma_{\textrm{el}}$.
The initial configuration of the box follows equilibrium multiplicities according to Maxwell-Boltzmann statistics (allowing for Poissonian fluctuations in each event).
The temperature of the system is calculated by fitting the momentum distribution of the pion momentum spectrum,
\begin{equation}
\frac{d^3N}{dp^3} \propto V e^{-\frac{\sqrt{m^2 + p^2}}{T}} \ .
\end{equation}
We considered the final temperature of the system to be equivalent to the temperature extracted from the pion spectrum because they are the most abundant stable particles in all our systems. For boxes where the only interaction between two particles has a constant isotropic cross section we took the initialization temperature as the final value of the system.
For a discrete set of time steps the correlation function can be calculated with the so called autocorrelation function
\begin{equation}
C(t) = \langle j_Q^x(t) j_Q^x(0) \rangle = \frac{1}{K - t} \sum_{s=0}^{K-t} j_Q^x(s\Delta t)j_Q^x(s\Delta t + t) \ ,
\end{equation}
where $K$ is the total number of time steps and $\Delta t$ the time interval between each time step. For every calculation we choose $\Delta t = 0.05$ fm. The total time elapsed 
for equilibration, after which we compute the transport coefficients, is chosen between $500$ fm and $900$ fm.
For a discrete case the electric current can be computed as,
\be \label{eq:current} j^x_Q (t)=\frac{1}{V} \sum_{i=1}^N q_i \frac{p_i^x (t)}{p_i^0(t)} \ , \ee
where $N$ is the number of particles in the system, and $p^0_i(t)=\sqrt{p^2_i(t)+m_i^2}$ the energy of the $i$th particle. The full phase space information of every particle in every time step is given by the SMASH output.
As previously discussed in \cite{Rose:2017ntg} there are different ways to extract the transport coefficient from the correlation function. In this work we choose to fit the average correlation 
function [see Eq.~(\ref{eq:expansatz})] and compute the electrical conductivity as described in Eq.~(\ref{eq:conducsimp}) with the intercept $C(0)$ and inverse slope parameter $\tau_Q$ extracted from the fit.
The number of events is chosen to be $\sim 600$ events per temperature point. The relative error is calculated for every time interval. With increasing time the error grows because there is only a finite set of time steps available to calculate the correlation function. For larger times the shape of the correlation function differs from an exponential shape and it is dominated by noise.
Therefore, it is important to determine a criterion up to which point one performs the fit. We chose this value to be $\sim 2\%$ of the relative error of the average correlation function.
To estimate the error of the final value of the electrical conductivity, we propagate the statistical uncertainties of $T$, $\tau_Q$ and $C(0)$. The volume of the box does not have an error. The uncertainty of the fit parameter $\tau_Q$ is determined by changing the fit range from $t = 2$ fm to the maximum fit range and computing the difference between the maximum and minimum value of $\tau_Q$ within this range. Therefore, the uncertainty of changing the cutoff criterion is taken into account.
The uncertainty of the intercept value $C(0)$ is chosen to be the error of the correlation function at $t = 0$ fm itself.
To extract $T$, we first compute the mean temperature of each individual time step over all events, and then average over time steps. The error of the temperature is computed from the mean squared uncertainty of averaging over all time steps.

We conclude this section by comparing our methodology to other similar approaches. In our previous work~\cite{Rose:2017bjz} we focus on the shear viscosity coefficient, using an analogous Green-Kubo approach. The work~\cite{Muronga:2003tb} follows similar lines in the context of the UrQMD transport code. In that work the shear-shear correlation function is also assumed to have an exponential decay. The values of the relaxation time and the shear viscosity itself are around a factor of 2 less than ours in~\cite{Rose:2017bjz}. However, the UrQMD model used in that work was revisited and upgraded in~\cite{Demir:2008tr}, where much closer values to our $\eta$ and $\eta/s$ are obtained for the whole temperature range. While the model used is different, in the two cases $\eta/s$ reaches a plateau at higher temperatures due to the presence of resonances whose lifetime is comparable to the mean-free time. 
Concerning the electrical conductivity, some previous references use similar Green-Kubo approach~\cite{Greif:2014oia,Puglisi:2014sha}. However, they study a partonic medium at temperatures higher to those considered here. Despite this difference, our expressions for $\tau_Q,C(0)$ and $\sigma_{\textrm{el}}$ are fully consistent with those presented in~\cite{Greif:2014oia} after the zero mass limit is taken. The same methodology as ours is used in~\cite{Puglisi:2014sha}. In these works, a (partonic) system of particles interacting via constant cross section has also been considered. In this case there is no difference with respect to our approach and the results are coincident after making the appropriate modifications of masses and cross sections. More about this will be commented on at the end of Sec.~\ref{sec:massless}.

\section{Test cases of simple systems~\label{sec:4}}

We present a set of results used to determine the validity of our approach. For this purpose, we use simple systems which can be tested against analytical or well-known results.

\subsection{Massive gas with constant cross section}

  To begin with, we simulate a gas of massive particles interacting via $2 \rightarrow 2$ collisions with a constant isotropic cross section of $\sigma=30$ mb. We consider three types of particles with electric charges $q_i=\{+1,-1,0\}$, all of them with equal mass $m_i=138$ MeV and equal densities. This system can be thought as a model for a pion gas interacting as hard spheres.

\begin{figure}
  \includegraphics[scale=0.4]{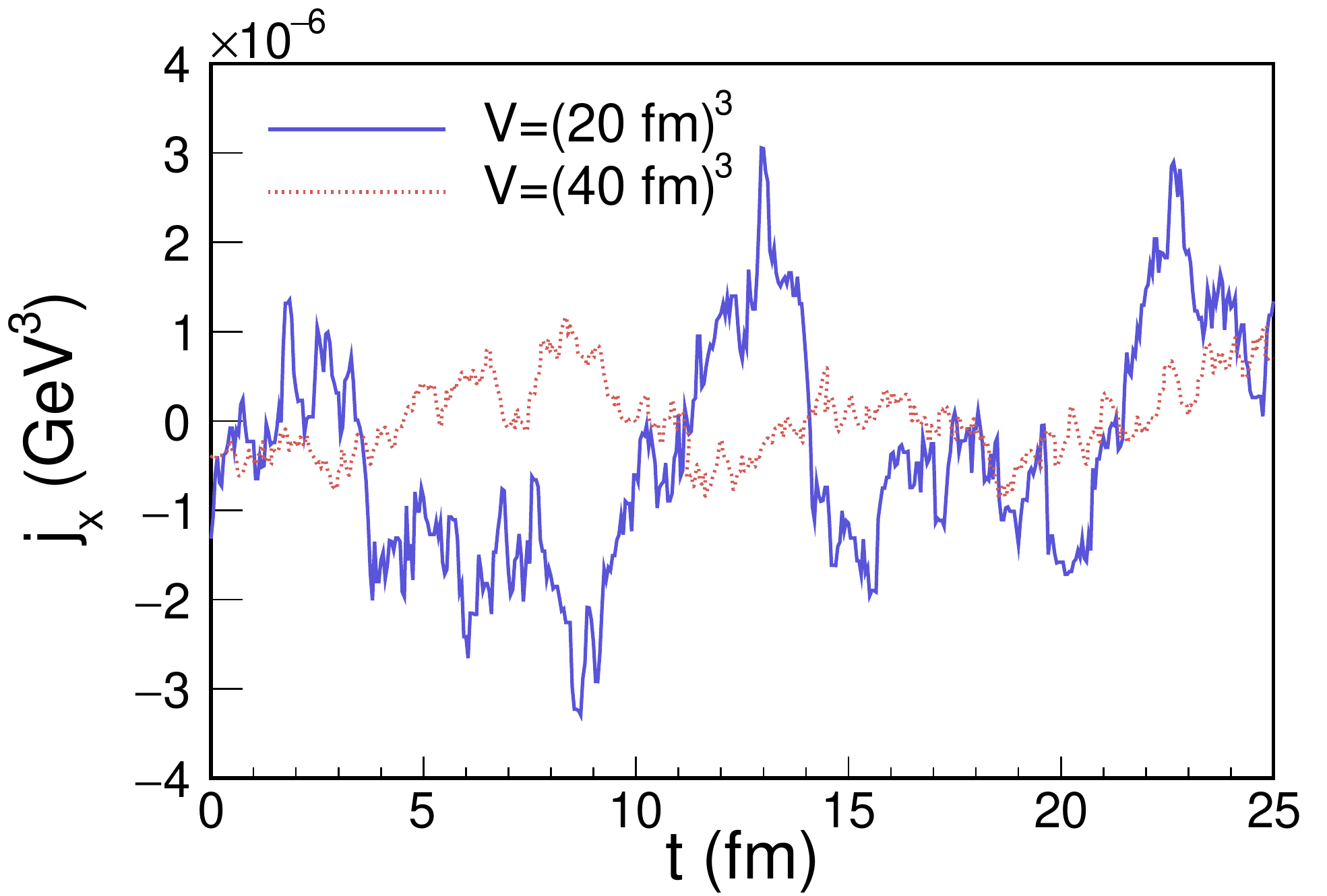}
   \includegraphics[scale=0.4]{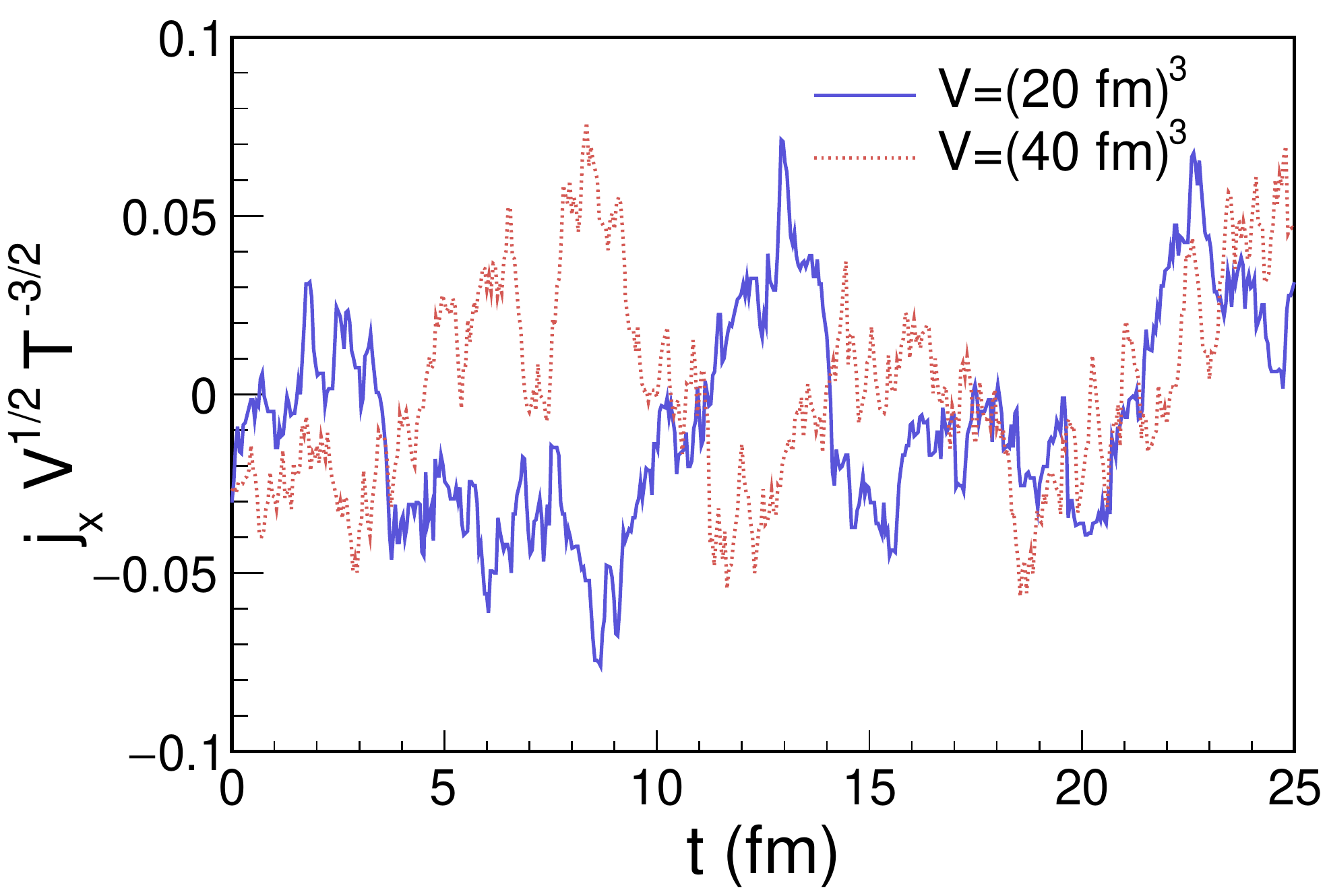}
  \caption{Left panel: Individual realizations for different box volumes of the electric current in the $x$ direction of a gas of massive particles interacting via a constant cross section. The temperature is $T=125$ MeV and the
  cross section $\sigma=30$ mb. Notice that the external electric field is zero. Right panel: Electric currents of the left panel normalized by $\sqrt{V}/T^{3/2}$ to get a volume-independent
  adimensional number.}
  \label{fig:noise}
\end{figure}

  In Fig.~\ref{fig:noise} we plot the $x$ component of the electric current, which can be computed as described in Eq.~(\ref{eq:current}). The system is kept at equilibrium with a temperature of $T=125$ MeV.

  In the absence of any external electrical field, the electric current is fluctuating around zero. Not only the time average of the electric current is zero, but the event average also vanishes.
Notice that we use two different box volumes: (20 fm)$^3$ (blue solid line) and (40 fm)$^3$ (red dotted line) to show that the strength of the fluctuations is $\sim 1/\sqrt{V}$.

\begin{figure}
  \includegraphics[scale=0.4]{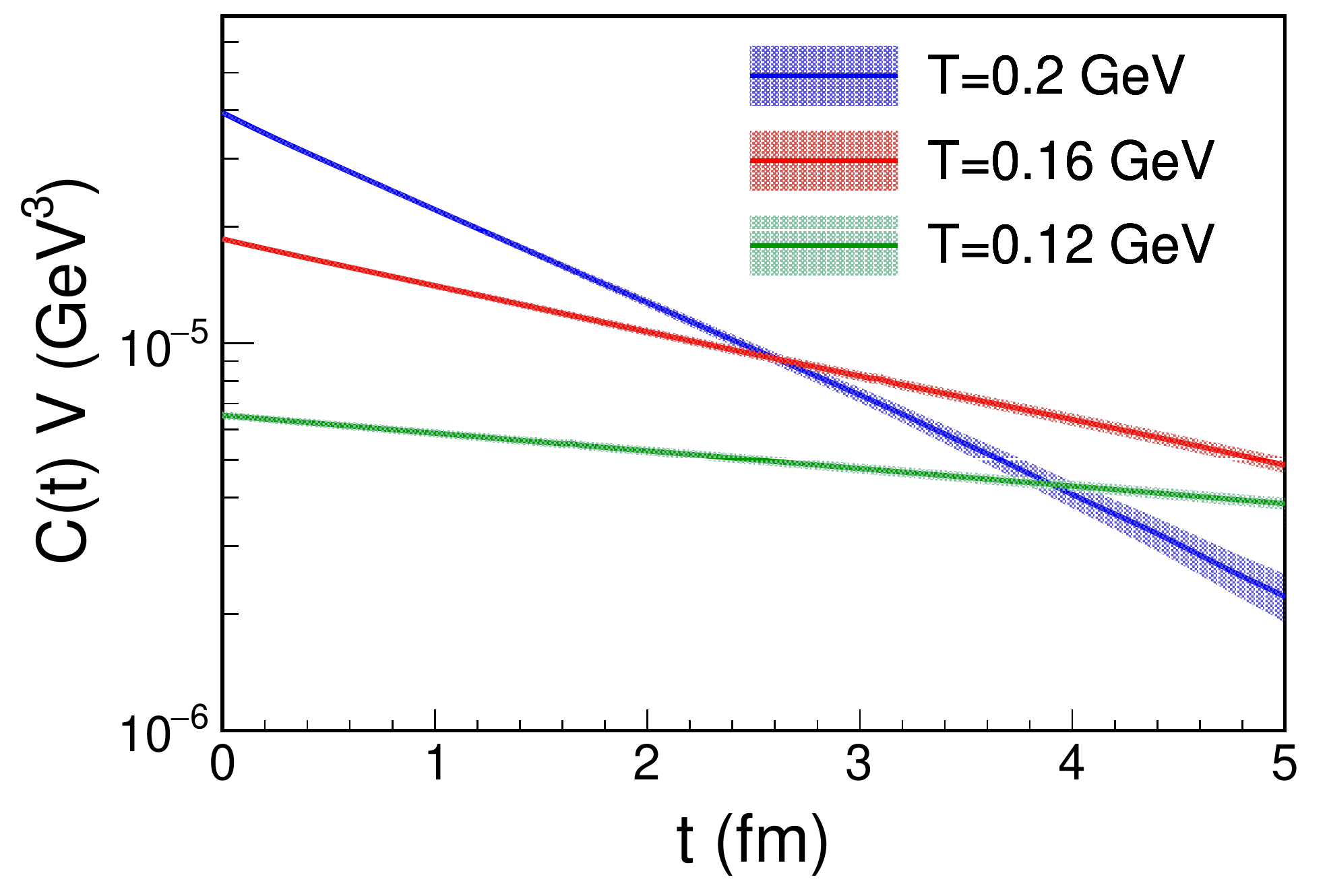}
  \caption{Correlation function times volume of a gas of massive particles at $T=0.12, 0.16$ and $0.2$ GeV interacting with a cross
  section of $\sigma=30$ mb.}
  \label{fig:corre}
\end{figure}

  In Fig.~\ref{fig:corre} we show the current-current correlation function as a function of time for a box at $T=0.12, 0.16$ and $0.2$ MeV, and $V=(30$ fm$)^{3}$. The exponential form assumed in Eq.~(\ref{eq:expo}) clearly fits the data at
short times. Due to the computation of only a finite number of time steps, the statistics of the correlation function $C(t)$ becomes unstable at larger $t$. We will assume that the exponential decay is valid for all times.

We perform an exponential fit as described in Section~\ref{sec:2.5}.
The slope of the fit (in semilogarithmic scale) is related to the relaxation rate of the current, whereas the intercept at the origin $C(0)$ is an equilibrium property, which measures the variance of the noise distribution.

It is possible to compute the value of the correlation function at the origin analytically~\cite{boon1980molecular,Greif:2014oia,Rose:2017bjz},
 \be \label{eq:preC0} C(0) = \sum_{a=1}^{N_s} \frac{g_a q^2_a e^2}{3V} \int \frac{d^3p}{(2\pi)^3} \left( \frac{p}{p_a^0} \right)^2 f_a (p)  , \ee
 where the sum runs over each species in the box with degeneracy $g_a$ and electric charge $q_a$.
 $f_a$ denotes the distribution function of the species $a$. Using Maxwell-Boltzmann statistics we have
  \be \label{eq:C0} C(0)V= \sum_{a=1}^{N_s} \frac{g_a q_a^2 e^2}{6\pi^2 }I(m_a,T,\mu_Q) \ , \ee
  with
  \be \label{eq:Ifunc} I(m_a,T,\mu_Q)=\int_0^\infty dp \frac{p^4}{m^2_a+p^2} e^{-\frac{\sqrt{m_a^2+p^2}-q_ae\mu_Q}{T}} \ , \ee
where $\mu_Q$ is the electric charge chemical potential.

  \begin{figure}
  \includegraphics[width=7cm,height=5cm]{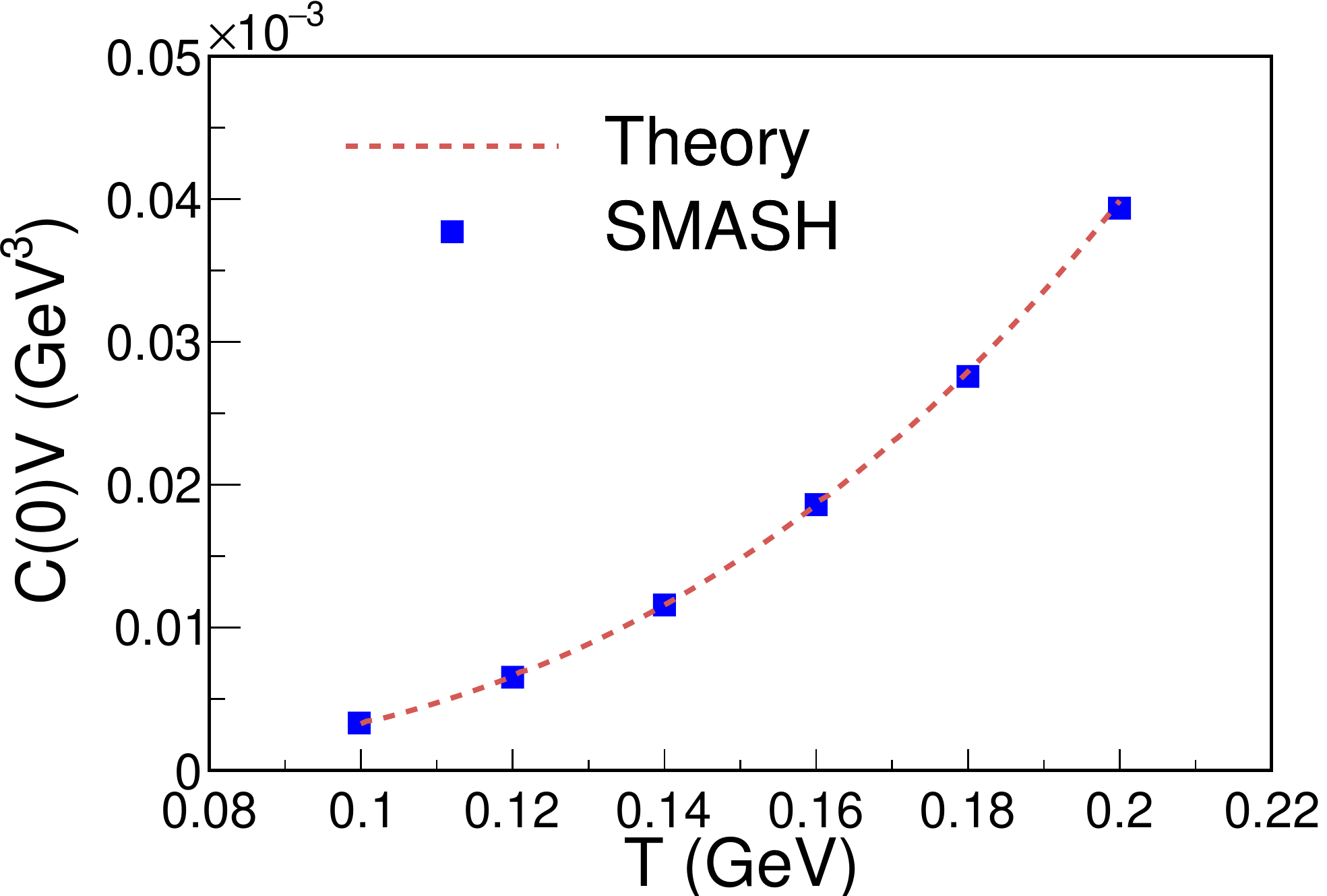}
  \includegraphics[width=7cm,height=5cm]{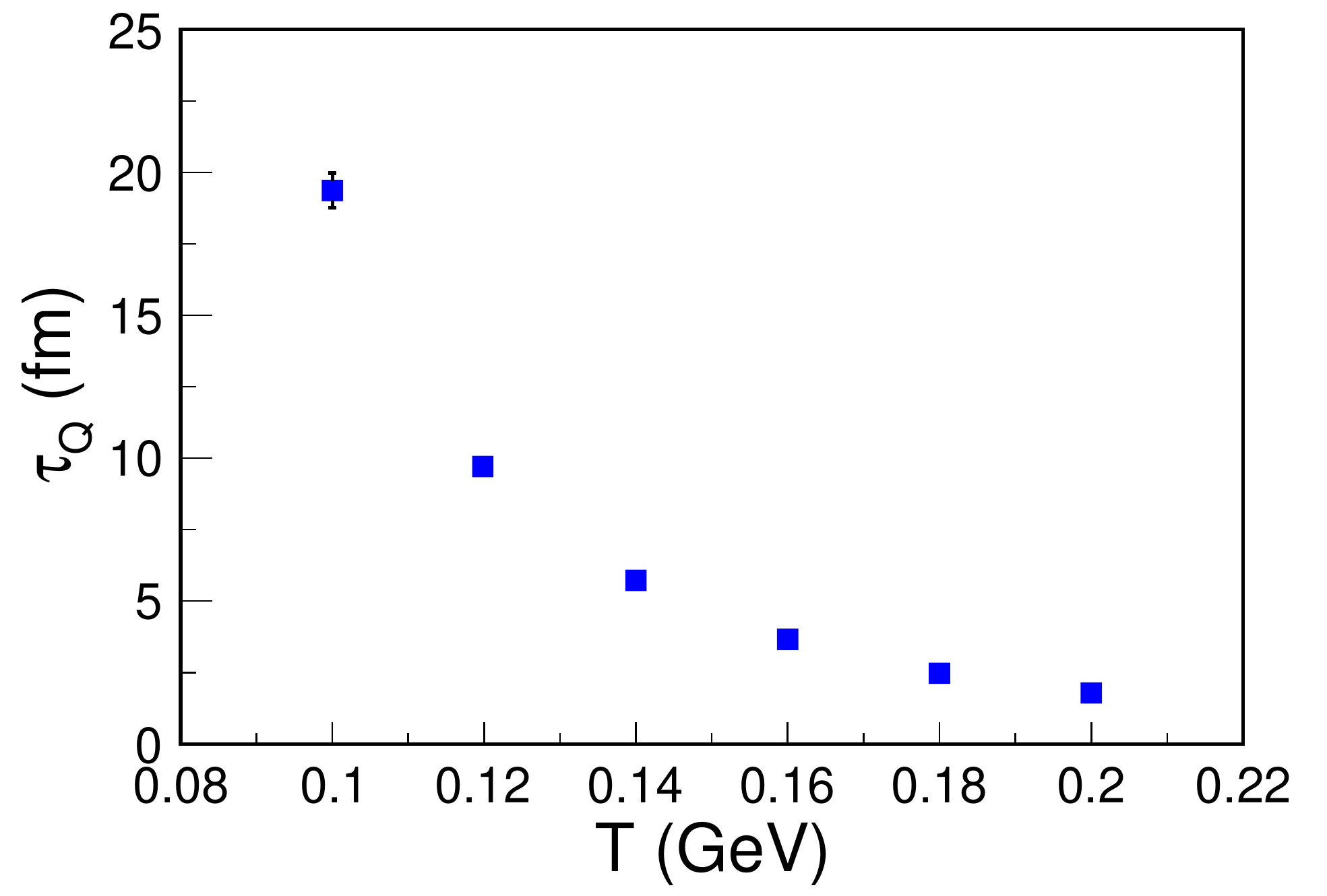}
  \caption{Left panel: Correlation function at $t=0$ (multiplied by the volume of the box) for a gas of massive particles as a function of the temperature. Dashed line: from Eq.~(\ref{eq:C0}), points: extraction
from SMASH. Right panel: Relaxation time for the same system as a function of the temperature.}
  \label{fig:C0massive}
\end{figure}

 In the left panel of Fig.~\ref{fig:C0massive} we observe that $C(0)V$ computed through Eq.~(\ref{eq:C0}) agrees very well with the measured values from SMASH for different temperatures of the box. In the right panel of the same 
 figure we present the relaxation time of the electric current, obtained after the exponential fit of the data. Notice that the relaxation time has a microscopical origin and it should be of the same order to the mean-free time. It decreases with temperature following the behavior of the mean-free time with 
 temperature as $1/(n\sigma) \simeq 1/(T^3 \sigma)$ (relativistic) or $1/ (\bar{v} n \sigma) \sim e^{m/T}/(mT^2 \sigma)$ (nonrelativistic). 
  
\begin{figure}
 \includegraphics[scale=0.4]{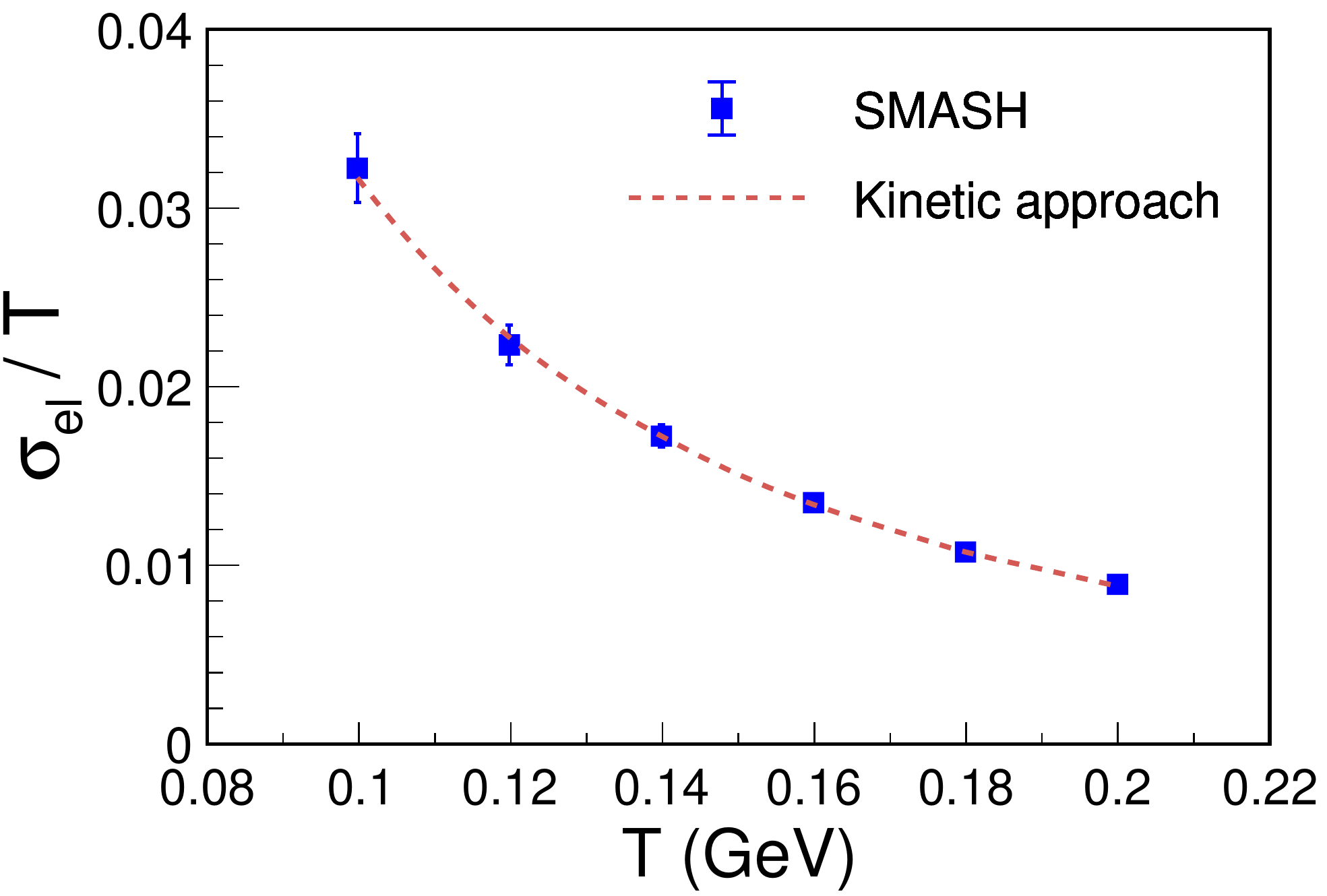}
\caption{Electric conductivity over temperature for a gas of massive particles interacting with constant cross section $30$ mb as a function of the temperature.
Symbols: Extraction from SMASH using Eq.~(\ref{eq:conducsimp}). Dashed line: results from the kinetic approach in Ref.~\cite{Greif:2016skc}.}
 \label{fig:sigmamassive}
\end{figure}

 
In Fig.~\ref{fig:sigmamassive} we present the electrical conductivity of the system for several temperatures. We compare the results using the Green-Kubo formula with the kinetic theory calculation 
presented in Ref.~\cite{Greif:2016skc}, where the Boltzmann equation is solved for the same system using a Chapman-Enskog-like expansion of the distribution function. The agreement is very good, as both
approaches solve the same Boltzmann equation with identical parameters. This constitutes an important numerical check of our algorithm.

\subsection{Massless pion gas and Drude formula\label{sec:massless}}

It is interesting to compare the massless case with the analogue to the Drude nonrelativistic estimate for the conductivity of massless particles.
This estimate in the nonrelativistic case reads~\cite{Drude}
\be \label{eq:Drudetau} \sigma_{\textrm{el}}^{\textrm{Drude,NR}}= \frac{q^2 e^2 n \tau}{m_q} \ . \ee
For the relativistic case, an analogous estimate has been obtained in~\cite{Greif:2014oia} in terms of the total cross section [$\tau=3/(2 n\sigma)$],
\be \sigma_{\textrm{el}}^{\textrm{Drude}}= \frac{1}{2} \frac{\sum_a q_a^2 e^2 n_a}{\sum_a n_a \sigma T} \ , \ee
where the sums run over the different species that compose the gas.
  
In our box calculation the density of each species is computed by counting the number of particles $n_a=N_a/V$, and the temperature is extracted by fitting the multiplicity as explained in the previous section.
The massless case is a bit more technical to handle due to the numerical treatment of massless particles. However, the analysis is even simpler as the temperature fixes the only running scale (cross section is kept constant).
  
For one massless species of charge $q$ the correlation function at the origin~(\ref{eq:C0}) reduces to
 \be \label{eq:C0massless} C(0) = \frac{g q^2 e^2}{3V} \int_0^\infty \frac{d^3p}{(2\pi)^3} e^{-\frac{p-qe\mu_Q}{T}} = \frac{(qe)^2 n}{3 V} = \frac{qe n_Q}{3V} \ , \ee
 where $n$ and $n_Q=qen$ are the particle density and the charge density, respectively. This equation has been previously obtained in~\cite{Greif:2014oia}, which corresponds to the limit $\lim_{m \rightarrow 0} I(m,T,0)=2T^3$.

     \begin{figure}
  \includegraphics[width=7cm,height=5cm]{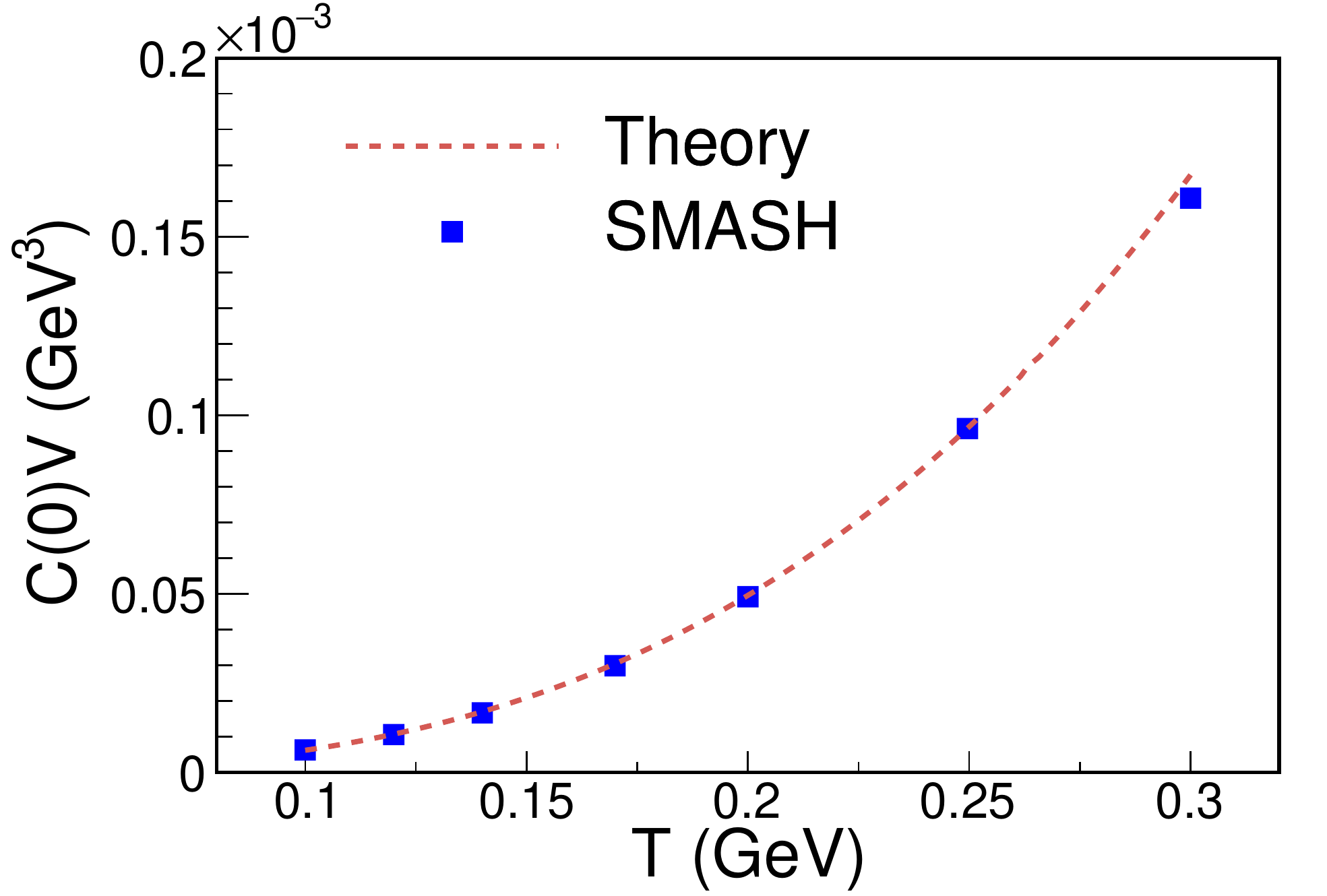}
  \includegraphics[width=7cm,height=5cm]{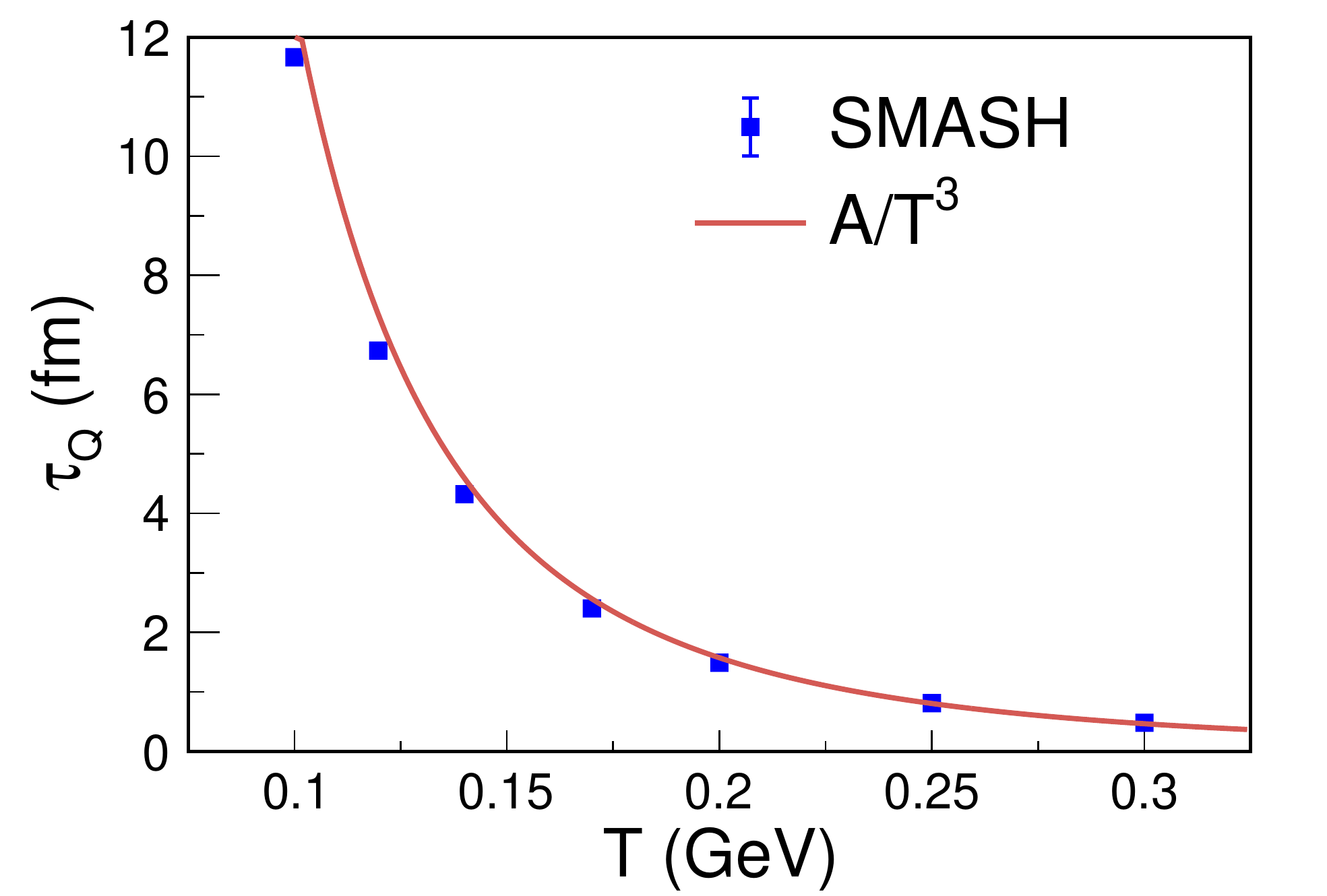}
  \caption{Left panel: Correlation function at $t=0$ multiplied by the volume of the box for a gas of massless pions as a function of the temperature. Dashed line: from Eq.~(\ref{eq:C0massless}). 
  Points: Extraction from SMASH. Right panel: Relaxation time as a function of the temperature.Points: Extraction from SMASH. Solid line: $A/T^{3}$ with $A=\pi^2/6$ fm$^{-2}$.}
  \label{fig:C0massless}
\end{figure}

For a massless gas with 3 charged species $q_a=\{-1,0,+1\}$ we show in the left panel of Fig.~\ref{fig:C0massless} $C(0)V$ together with the analytical estimate. In the right panel 
we show the relaxation time as a function of the temperature which follows the expected $\tau_Q=A/T^3$ behavior. If we identify $\tau_Q$ with $\tau=3/(2n\sigma)$ in the Drude formula we can even estimate the coefficient $A$. Using $n=gT^3/\pi^2$ for a massless Boltzmann gas, with $g=3$ we obtain $A=\pi^2/(2\sigma)=\pi^2/6$ fm$^{-2}$. We plot $A/T^3$ together with our data to see the very good matching between the two.
  
\begin{figure}
\includegraphics[scale=0.4]{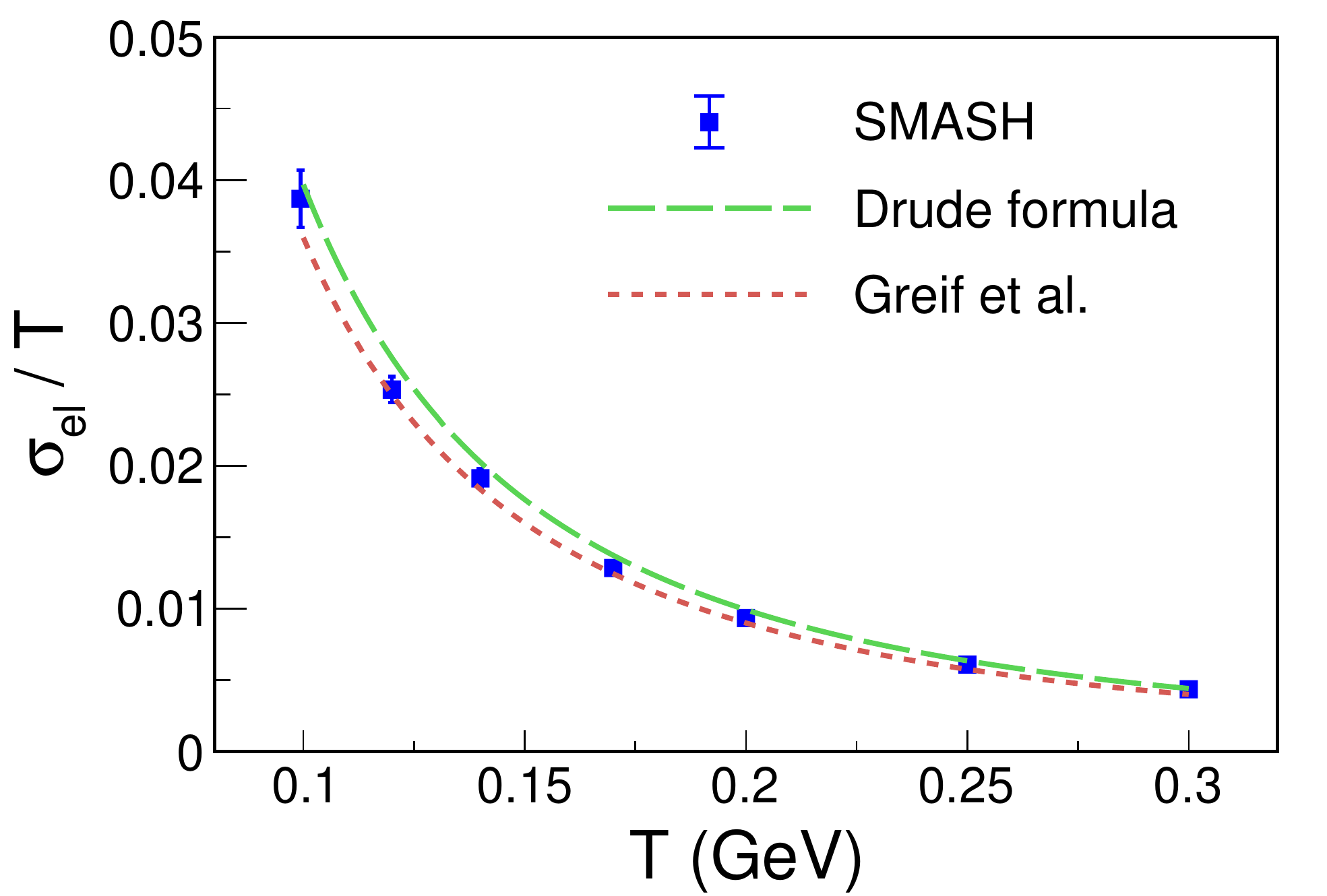}
\caption{Electric conductivity over temperature for a gas of massless pions interacting with constant cross section $30$ mb as a function of
the temperature. Points: Extraction from SMASH. Long-dashed line: results from the Drude formula (\ref{eq:drude}). Short-dashed line: estimate (\ref{eq:greif}) from Ref.~\cite{Greif:2014oia}. }
\label{fig:sigmamassless}
\end{figure}

The electrical conductivity of the system is plotted in Fig.~\ref{fig:sigmamassless} together with the Drude estimate for massless particles. For the conditions of our box ($\sigma=30$ mb) this estimate reads
  \be \label{eq:drude}  \sigma_{\textrm{el}}^{\textrm{Drude}} = \frac{1}{2} \frac{2e^2}{3 \sigma T} = \frac{3.9686 \cdot 10^{-4}}{T} (\textrm{GeV}^2) \ , \ee
where we used $e^2=4\pi \alpha=4\pi/137$.
  
We observe a slight underestimation of our data compared to the Drude formula, although there is a general agreement. A more precise approximation to the electrical conductivity can be found in Ref.~\cite{Greif:2016skc} where an analytical expression is obtained using kinetic theory. In that reference the estimate is shown for a quark-gluon plasma gas interacting with constant cross section.
It is easy to modify the prefactor $\sum_a q_a^2 g_a/\sum_a g_a$ to account for the correct degrees of freedom of our massless box.
After dividing over the QGP prefactor $2/13$ and multiplying by the prefactor in our box $2/3$, we simply have to rescale the result to account for the different cross section (30 mb instead of 3 mb).
One obtains
\be \label{eq:greif} \sigma_{\textrm{el}} = \frac{2/3}{2/13} \times \frac{3}{30} \times \frac{0.000832737}{T} = \frac{3.5967 \cdot 10^{-4}}{T} (\textrm{GeV}^2) \ . \ee  

This result is approximately 10\% smaller than the Drude estimate. While the Drude estimate is equivalent to a relaxation-time approximation [see Eq.~(\ref{eq:Drudetau})], the kinetic approach contains the whole collision integral of the Boltzmann equation, and it represents a better approximation to the SMASH results for the same system.  The comparison is excellent as shown in Fig.~\ref{fig:sigmamassless}.

In addition, we have checked that the electrical conductivity is independent of the density of charge carriers (as evident from Drude's estimate). As an additional check we computed the electrical conductivity of a gas without neutral particles, so that the density of scatterers is reduced a factor 2/3. The obtained electrical conductivity is, as expected, a factor 2/3 smaller than the previous case.

\subsection{Mixture of massive hadrons with constant cross section}

To increase the complexity of the hadron plasma we simulate the dynamics of a mixture of particles interacting via constant isotropic cross sections. We implement massive pions, kaons and nucleons interacting among each other 
with a constant cross section of $30$ mb. Although the precise value of the cross section is not very relevant now, it is important to mention that all scatterings occur via elastic (and local) reactions.

\begin{figure}
  \includegraphics[scale=0.4]{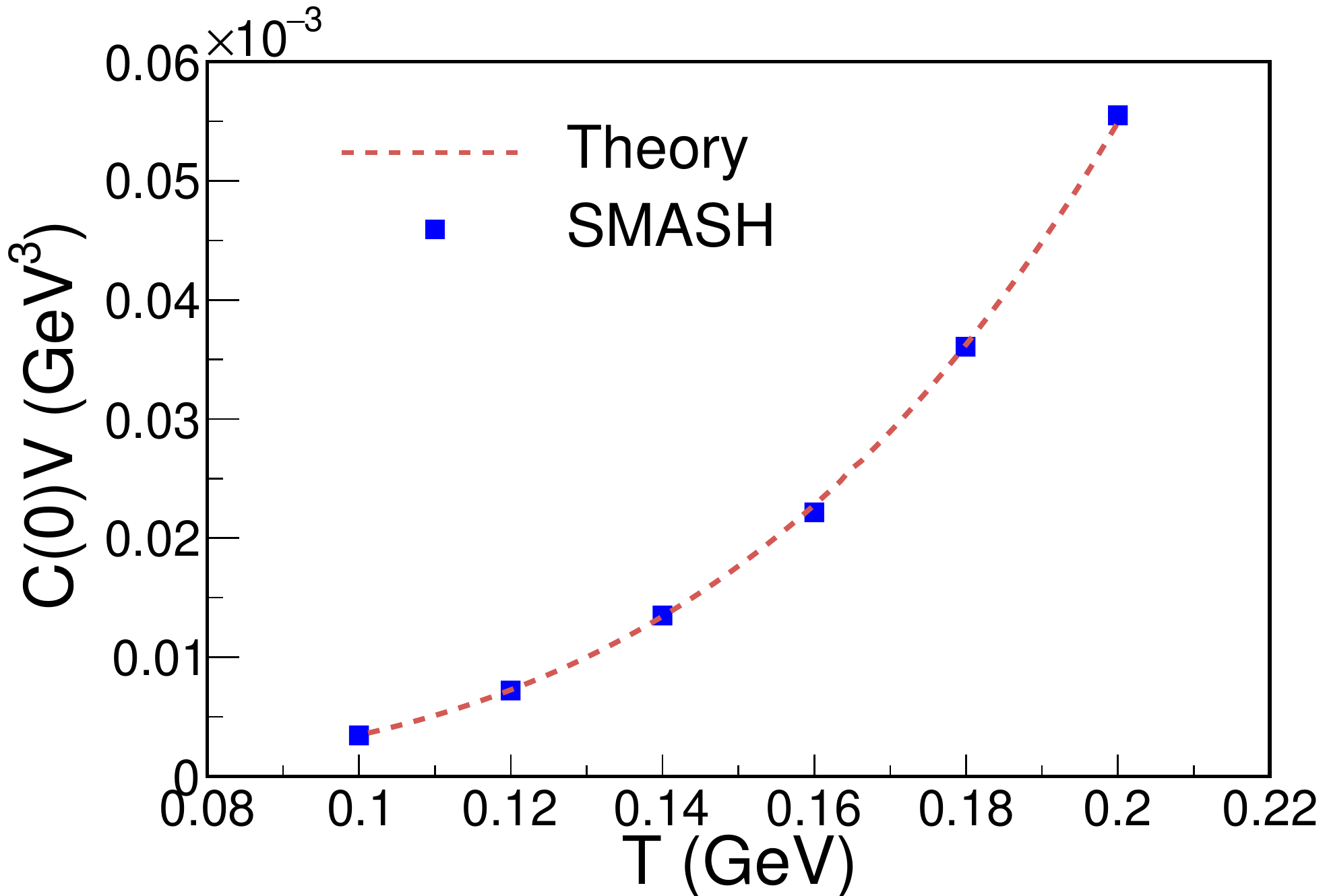}
  \includegraphics[scale=0.4]{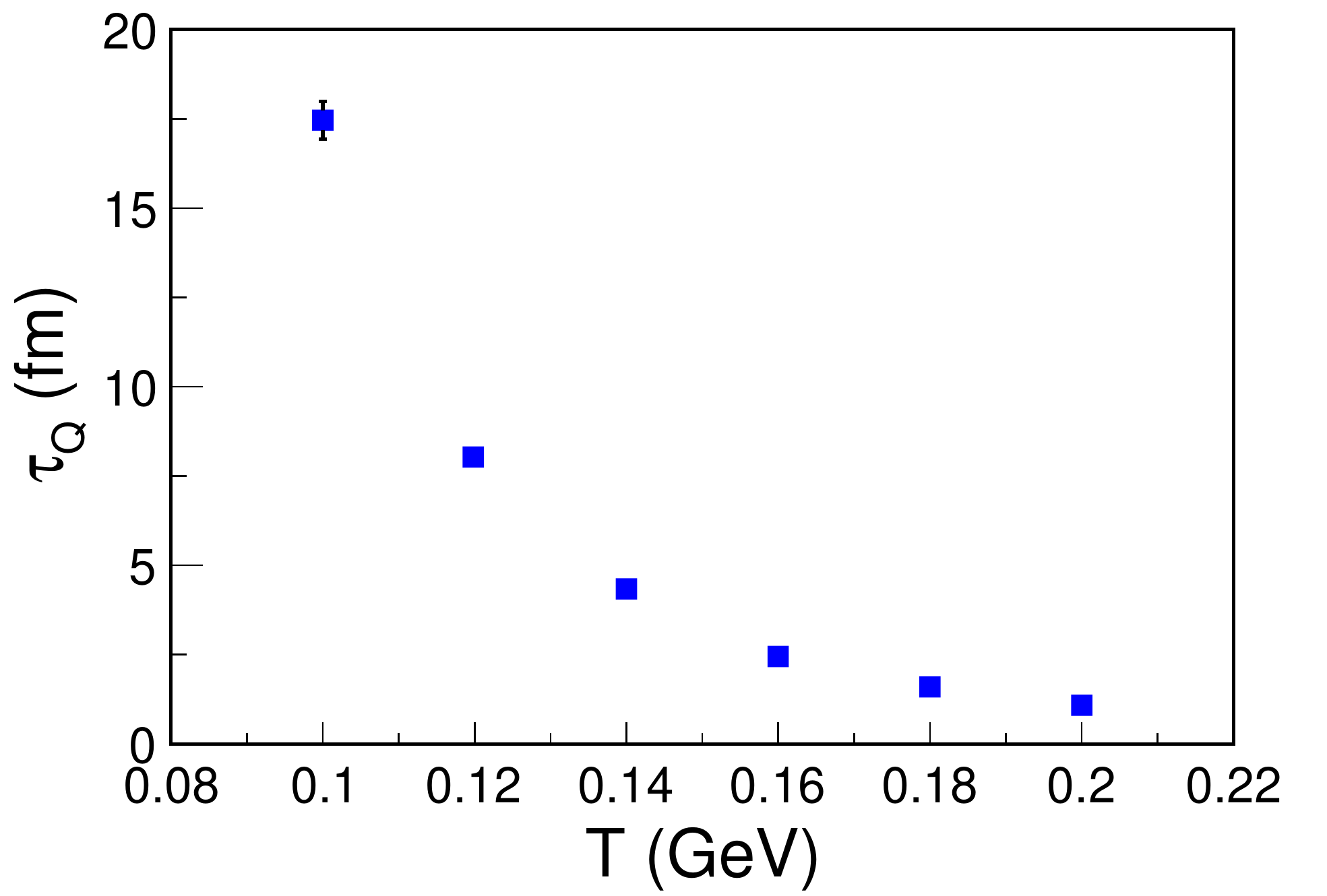}
  \caption{Left panel: Correlation function at $t=0$ for a gas of massive pions, kaons and nucleons as a function of the temperature. Interaction among particles are elastic with a cross section of $30$ mb.
  Dashed line: from Eq.~(\ref{eq:C0massless}). Points: Extraction from SMASH. Right panel: Relaxation time as a function of the temperature.}
  \label{fig:C0mixture}
\end{figure}

As seen in the left panel of Fig.~\ref{fig:C0mixture} the extracted value of $C(0)V$ is in excellent agreement with the theoretical prediction (now with 3 different species). In the right panel 
we show the value of the relaxation time of the electric current as a function of the temperature. 

\begin{figure}
  \includegraphics[scale=0.4]{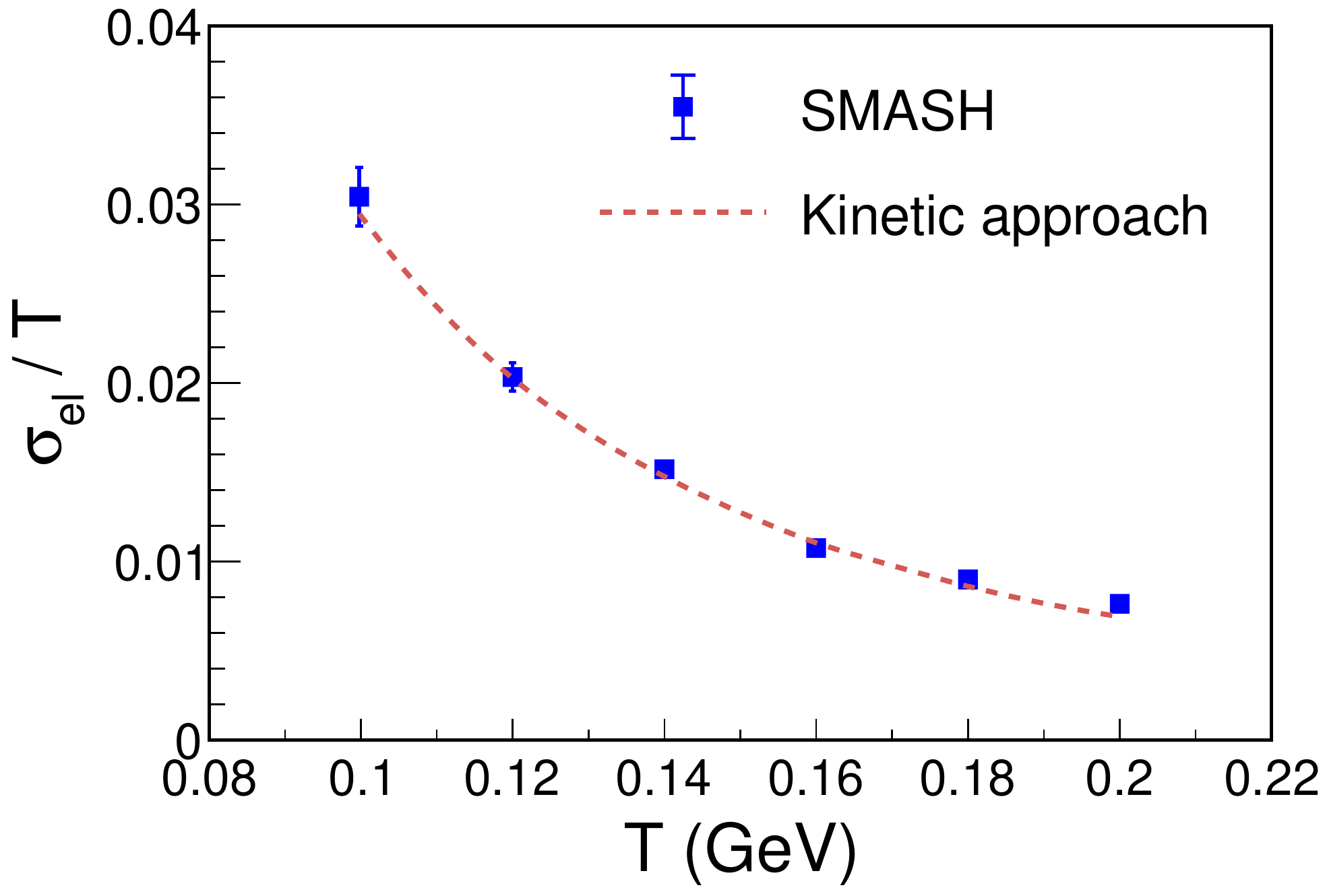}
  \caption{Electric conductivity over temperature for a gas of massive pions, kaons and nucleons interacting with constant cross section $30$ mb as a function of the temperature. Points: Extraction from SMASH.
  Dashed line: estimate using the kinetic approach in Ref.~\cite{Greif:2016skc}.}
  \label{fig:sigmamixture}
\end{figure}

In Fig.~\ref{fig:sigmamixture} we observe that the Green-Kubo method works very well for the hadron mixture too. We compare with the kinetic theory calculation of Ref.~\cite{Greif:2016skc} for the same system. 

\section{$2 \rightarrow 2$ collisions vs explicit resonance formation/decay~\label{sec:5}}

 In this section we apply the SMASH code with more realistic interactions. The scattering is produced by formation and decay of resonances i.e. through $2 \leftrightarrow 1$ reactions. For a process with two pions in the initial and final states, the two mechanisms: 1) $\rho$ creation in the medium, propagation, and eventual decay and 2) energy dependent elastic cross section, describe well the interaction between them. However, as recently pointed out in Ref.~\cite{Rose:2017bjz} for the shear viscosity, the two have a quite different influence on transport coefficients.

  The idea put forward in~\cite{Rose:2017bjz} is that the redistribution of a conserved quantity allowing for the equilibration of a local perturbation, might be effectively blocked within
times of the order of the resonance lifetime. At low temperatures the mean-free time is large enough (compared to the resonance lifetime) and the transport coefficients should be insensitive
to the particular form of the interaction (elastic or inelastic). However at higher temperatures, where the mean-free time decreases, finite lifetimes slow down the equilibration process, thus
increasing the relaxation time and therefore, the transport coefficient. This effect is relevant for the shear viscosity case~\cite{Rose:2017bjz} as the absolute difference between the two scenarios in the shear viscosity is 50 \% at the highest temperatures. 

\begin{figure}
  \includegraphics[scale=0.4]{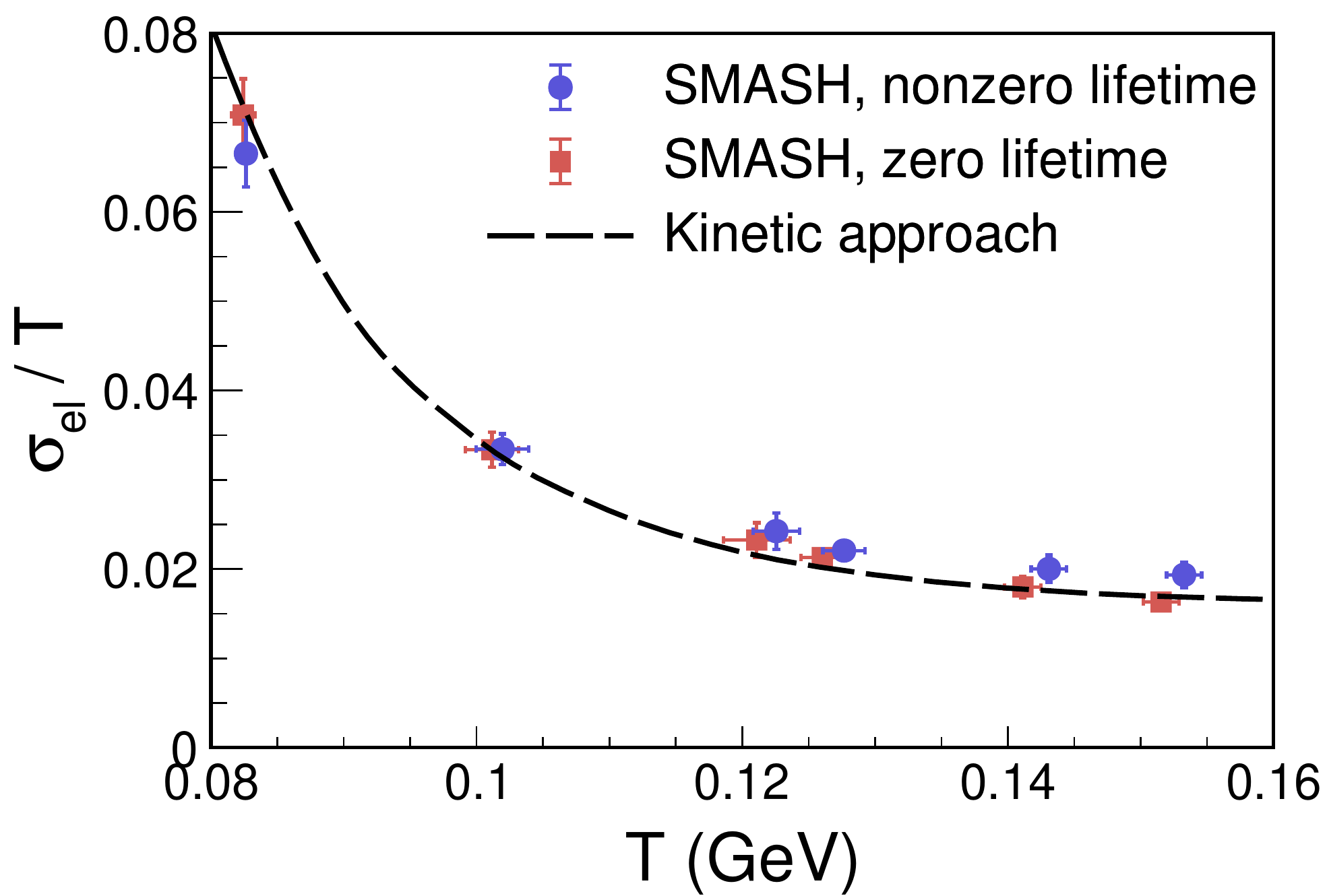}
 \includegraphics[scale=0.4]{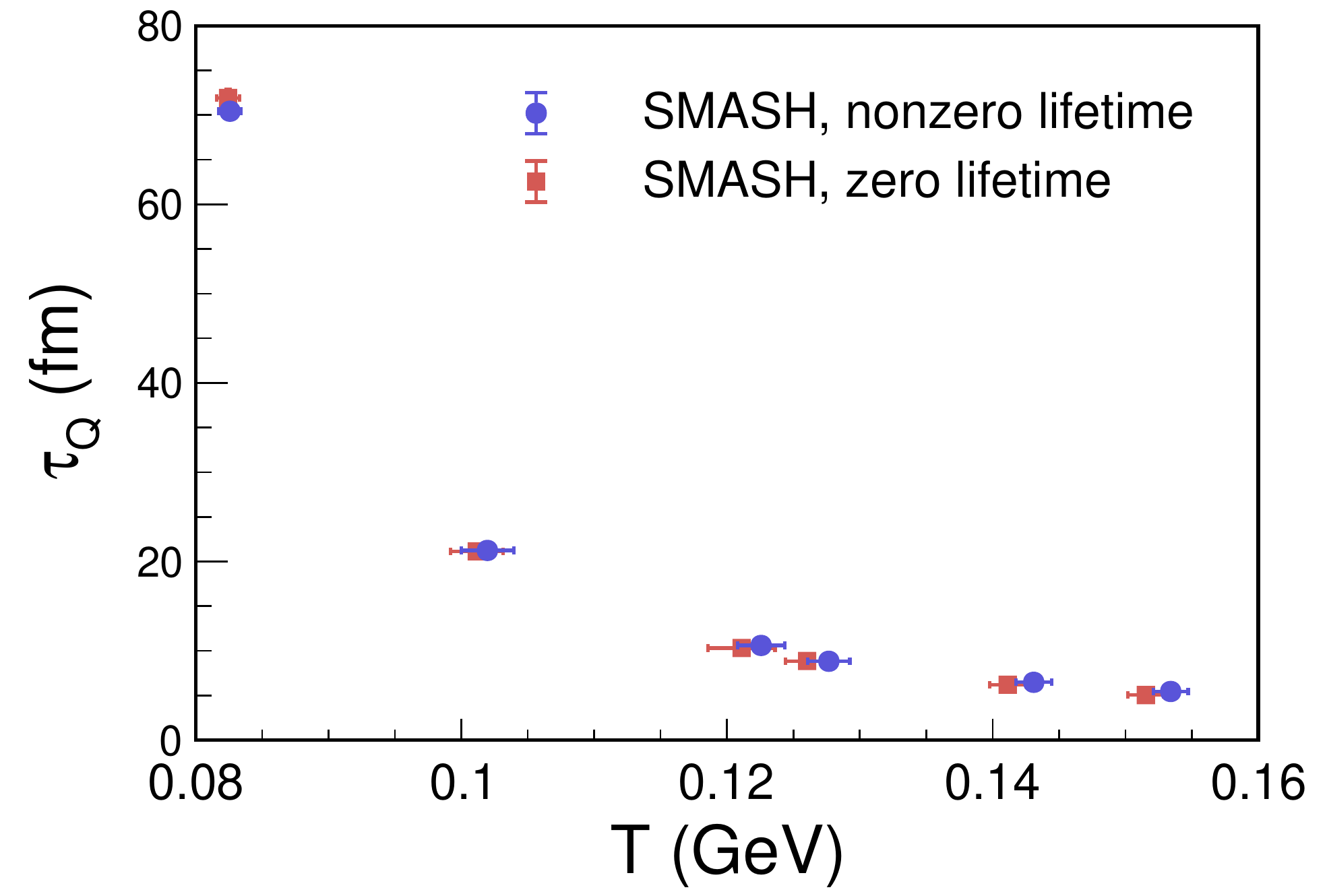}
  \caption{Left panel: Electric conductivity over temperature for a gas of pions and $\rho$ mesons. Open circles: SMASH result taking into account the physical lifetime of the $\rho$ meson. Squares: SMASH result when 
  the lifetime of $\rho$ meson is artificially set to zero.
 Dashed line: estimate from Ref.~\cite{Greif:2016skc} based on kinetic theory. Right panel: SMASH results for the relaxation time for the same system
  in the left panel.}
  \label{fig:sigmarhopi}
\end{figure}

  With this idea in mind we study a simple $\pi-\rho$ box and compute the electrical conductivity under the two descriptions. Our result is summarized in Fig.~\ref{fig:sigmarhopi}. 
  
  In circles we show the result for the electrical conductivity from SMASH where pions interact via the creation of a $\rho$ meson, which propagates without further interaction until 
it decays. Square symbols represent the results for which we have artificially set the $\rho$ meson lifetime to zero, in such a way that it is forced to decay immediately after its formation.
The latter calculation pretends to mimic a $2 \rightarrow 2$ contact interaction. As seen in Fig.~\ref{fig:sigmarhopi} the second approach is closer to the kinetic approach
of Ref.~\cite{Greif:2014oia}, which cannot capture lifetime effects, as it only contains local $2 \leftrightarrow 2$ processes in the collision integral.

  From this plot it is clear that, as opposed to the shear viscosity, the lifetimes play no relevant effect in the relaxation time, see right panel of Fig.~\ref{fig:sigmarhopi}. 
The electrical conductivity $\sigma_{\textrm{el}}$ presents a slight increase for the case with nonzero lifetime (no larger than 20 \% at $T\simeq 150$ MeV) but the effects can be almost entirely ascribed to 
the difference in the $C(0)$ value. 

Why does the lifetime not play a significant role in the electrical conductivity? We can understand the effect by looking at the realization of the transport process at the microscopical level. 
  
For the shear viscosity one needs to equilibrate the momentum excess of particles (with respect to fluid's cell velocity) between different regions. Collisions between particles isotropize the momentum distribution to reach 
equilibrium. When this mechanism is entirely produced via resonance formation, it is necessary to wait until the formed resonance decays for the total momentum to be distributed between the decay products. 
The real transport is only effective at the moment of the decay, but not before. If the lifetime is larger than the mean-free time, then, on average, it will dominate the value of the relaxation time $\tau_Q$. 

  Although the case of electric current seems to follow the same logic, there are some differences when electric charge enters into play. From all possible interaction processes, there are several cases in which the
equilibration (the disappearance of a fluctuation of the electric current) does not need to wait until the decay of the resonance. An initial fluctuation in the current might vanish at the interaction point, when the resonance 
is formed. As an example, one can think of a local electric current fluctuation due to a $\pi^+ \pi^-$ pair with an imbalanced total momentum. At the collision point a $\rho_0$ resonance with nonzero momentum is formed. 
One might think that the $\rho_0$ is required to decay into a pair with balanced momentum, to erase the electric current. However, the resonance is charge neutral and the initial fluctuation is already blurred at its formation, so the lifetime does not play any role in the relaxation time. Imagine now a local current created by a pair of a charge and a neutral pions, with a similar momentum but opposed in direction.
After they collide a nearly static charged $\rho$ meson is formed, which will live for some time. But again, the local current decreases close to zero at the formation time. 

In the hadronic gas, we have many pions of different charges creating local fluctuations and colliding in many different charge combinations. While the whole situation is much more complicated and difficult to analyze, we can convince ourselves by these examples, that the lifetime will play, at least, a minor role in $\tau_Q$ as compared to the shear relaxation time. And this is what we observe in our results.

Although of minor relevance for the electrical conductivity, we still want to stress that finite lifetimes might be of importance in the physics of equilibration, when their values become comparable to the 
typical mean-free time. In such a case the relaxation time cannot be traded simply by the mean-free time, as done in many phenomenological applications. These approaches miss the effect of time delay
of resonances, which becomes important for the extraction of transport coefficients, especially at higher temperatures. On the other hand, it is fair to say that pure elastic collisions should also contribute
to the dynamics in addition to the resonance interaction, for example to describe repulsive interactions. As seen in~\cite{Rose:2017bjz}, the addition of the elastic cross section will decrease the relaxation
time at high temperatures, as these processes are not affected by finite lifetimes.

\section{Results for a hadron gas~\label{sec:6}}

   Finally we apply SMASH to compute the electrical conductivity of a hadron gas modeled as a mixture of pions, kaons and nucleons, with the most prominent resonances ($\rho$ meson, $K^*$ and $\Delta$, cf. Table~\ref{tab}).
This system represents a relatively simple and reliable model for a physical hadron gas in equilibrium at temperatures below the phase transition temperature $T_c$.

\begin{figure}[ht]
\includegraphics[scale=0.4]{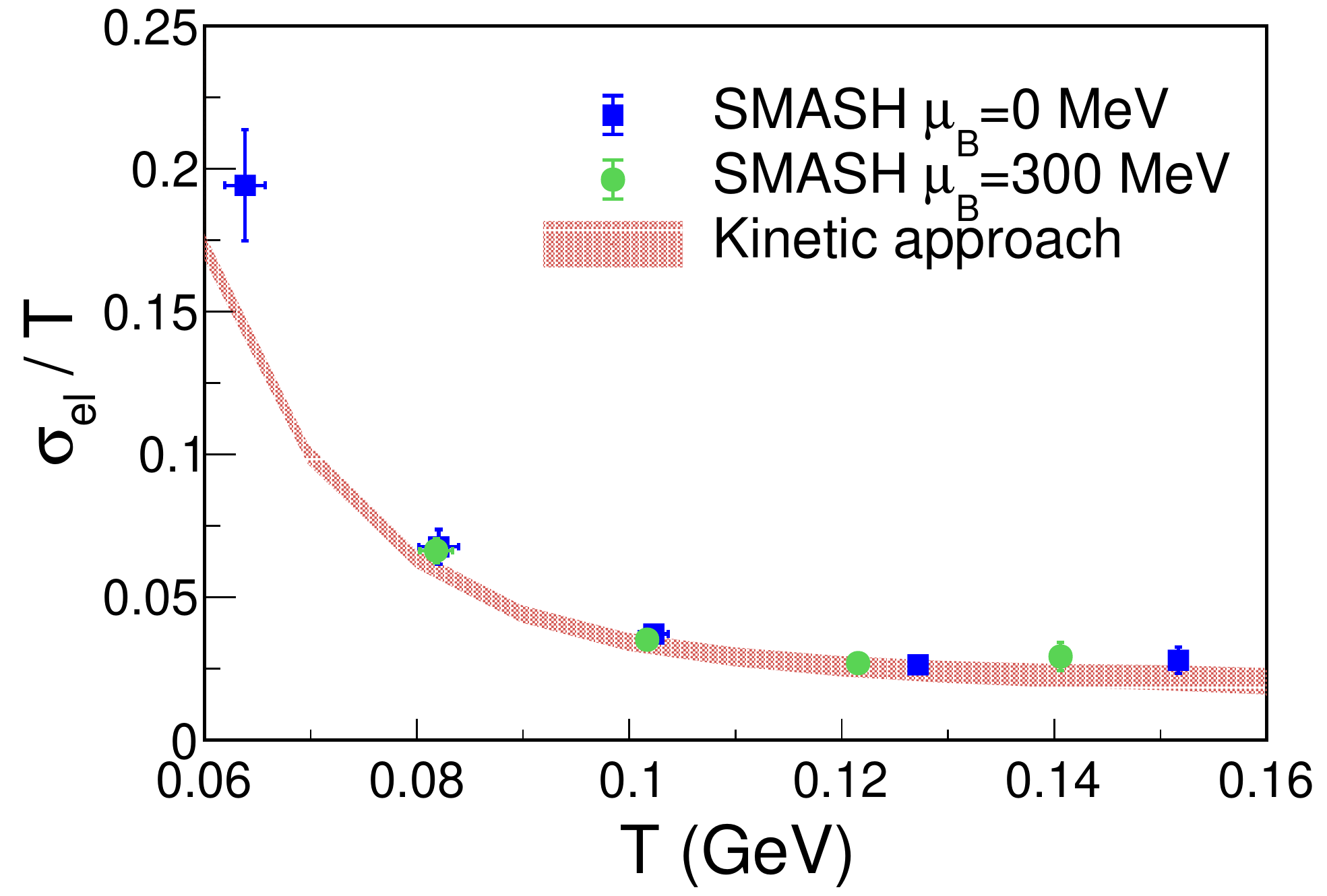}
\caption{Electric conductivity over temperature for a hadron gas containing $\pi,\rho,K,K^*,N$ and $\Delta$ baryons as a function of the temperature. Symbols show SMASH results for 2 different baryochemical potentials
(squares $\mu_B=0$  MeV, and circles $\mu_B=300$ MeV) and the kinetic approach ($\mu_B=0$ MeV) is an adaptation from Ref.~\cite{Greif:2016skc} with the cross sections used by SMASH.}
\label{fig:full}
\end{figure}

We show in Fig.~\ref{fig:full} the results from SMASH and compare them to those of Ref.~\cite{Greif:2016skc} at zero baryonic chemical potential. Indeed, we observe similarities in the shapes and value ranges of the two curves, although these are not perfect. This is however expected, as the two models are actually not complete mirrors of each other. In fact, the calculations from the kinetic approach include constant elastic cross sections between many of the possible particle pairs (see Table 1 in Ref.~\cite{Greif:2016skc} for details), whereas SMASH neglects some of these (e.g. elastic $\pi^+ \pi^+$ or $\pi^- \pi^-$collisions) to focus on resonance formation and propagation. We would thus actually expect SMASH results to be slightly larger than those of the kinetic approach, but the fact that both calculations give similar results is an encouraging step toward theoretically constraining the electrical conductivity.

Figure~\ref{fig:full} also shows the values of the electrical conductivity for nonzero baryonic chemical potential $\mu_B$ as calculated in SMASH. Up to $\mu_B =300$ MeV, we observe no significant variation. This effect is similar to what was obtained previously in the case of the shear viscosity to entropy density ratio (Ref.~\cite{Rose:2017bjz}, Fig. 8). It can be similarly understood: such an increase in the baryonic chemical potential at these temperatures representing only a few per cent of the total baryons in the system,
with pions still largely dominating charged particle multiplicities, one should also not expect massive variations in transport coefficients for such systems.
While we tried to increase the baryochemical potential to study the effect of nucleons and $\Delta$ baryons, we observed that the exponential {\it ansatz} for $C(t)$ ceases to be valid, invalidating our assumptions and methodology for such densities. This is why we do not include results at larger $\mu_B$.

 The comparison with previous estimates can be seen Fig.~6 of the recent Ref.~\cite{Ghosh:2018kst}. In that plot, our result (similar to {\it Greif et al.} line) is close to other approaches with slightly different interaction, e.g. the sigma model with and without medium modifications via the relaxation time
approximation of the Boltzmann equation, and the unitarized chiral perturbation theory via Green-Kubo technique~\cite{FernandezFraile:2005ka}. The differences among all these approaches in $\sigma_{\textrm{el}}/T$ is $\sim 27 \%$
at temperatures of $T=100$ MeV, and increase up to $\sim 40 \%$ for $T=160$ MeV.

\section{Conclusions~\label{sec:7}}

In this work we have applied the SMASH transport code in a ``box configuration'' at equilibrium to extract the transport coefficients for an interacting charged hadron gas. In particular, we have simulated
different systems at finite temperatures and densities, and computed the relaxation time of the electric current and the electrical conductivity. Based on the relativistic theory of hydrodynamic fluctuations, in
connection with Green-Kubo formulas, these coefficients have been extracted as functions of temperature and baryochemical potential, via the current-current correlation function.

In the hydrodynamic description of fluid systems, these parameters can be used as inputs for the hydrodynamical evolution of the hadronic matter. As a very simple example in 1+1 dimensions, this has been illustrated in App.~\ref{app}, for 
an initial Dirac delta profile of the charge density. In a full hydrodynamic code, a numerical scheme must be used to solve the whole set of coupled hydrodynamic equations of motion. In that case, these coefficients
(plus others, like the shear viscosity~\cite{Rose:2017bjz}) can be used as inputs in a similar way.

The results obtained for the electrical conductivity are consistent with previous estimates in the literature for similar systems. In particular, our result for the pion gas (with resonant interaction via $\rho$ meson) is
in total agreement with the kinetic theory calculation of~\cite{Greif:2016skc} when similar interactions are incorporated.

At high temperatures, the result for the full hadron box are expected to match the lattice-QCD calculations close to $T_c$, the crossover temperature. However, in that region, lattice-QCD results deviate significantly
from the hadronic approaches, cf.~\cite{Greif:2016skc}. For example, the value of $\sigma_{\textrm{el}}/T$ at $T=150$ MeV is close to $\sigma_{\textrm{el}}/T=0.003$, according to the results in Ref.~\cite{Amato:2013naa}.
Above $T_c$, lattice-QCD values rapidly increase to values, which are close to our estimates. Therefore, one needs to understand the reason of the rapid decrease of the electrical conductivity in lattice-QCD results when crossing $T_c$ from above.
For example, in~\cite{Brandt:2015aqk} a value of $\sigma_{\textrm{el}}/T\sim 0.04$ is quoted for $T=0.169$ GeV.
A more recent value right above $T_c$ is computed in Ref.~\cite{Ding:2016hua} giving a band between $\sigma_{\textrm{el}}/T \sim (0.010-0.036)$. These estimates from lattice QCD would match well with our result at $T<160$ MeV for the hadron gas (see also e.g.~\cite{Greif:2016skc}). Even in this case one should note that in the calculation of~\cite{Brandt:2015aqk} the pion mass is twice the physical one, whereas in~\cite{Ding:2016hua} there are no dynamical quarks (quench approximation). Therefore, any comparison has to be done with caution.

Finally, we comment on possible extensions of this work. A nonzero baryochemical potential is straightforward to implement in our scheme. We have presented some results at $\mu_B=300$ MeV, but they do not appreciably differ from the net baryonless case at $\mu_B=0$. The use of higher values of the baryochemical potential present a conflict in our approach: the exponential {\it ansatz} for the current-current correlation function ceases to be valid. We have already observed this effect in our simulations at high densities (not included in our analyses). A possible direction could be to explore denser systems beyond
the exponential {\it ansatz}. Incidentally, the numerical time integration of the current-current correlation function is difficult to perform due to the huge uncertainties at large times~\cite{Rose:2017ntg}. 

Finally, there are possible improvements to the results presented here. First, one can add more species to the final determination of the coefficients. A few more massive states will not contribute much at the temperatures considered here and $\mu_B=0$, but the addition of the full set of hadron states (resonance gas) might become important, especially close to $T_c$. The increase of the baryochemical potential (relevant for the approach to the possible critical region of QCD) will make baryons to dominate the dynamics, and contribute considerably to the transport coefficients.
But as mentioned already, this limit enters in conflict with the exponential {\it ansatz} used. The use of medium-modified interactions would also have influence on the transport coefficients. However, this would entail
a large modification in the structure of the SMASH code, and the expected variation of $\sigma_{\textrm{el}}/T$ is expected to be of the order of $\sim 20\%$~\cite{Ghosh:2018kst} for the top temperatures considered here (and less than $\sim 5 \%$ at $T=100$ MeV).

\acknowledgments

This work was made possible thanks to funding from the Helmholtz Young Investigator Group VH-NG-822  from  the  Helmholtz  Association  and  GSI,
and  supported  by  the  Helmholtz  International  Center for  the  Facility  for  Antiproton  and  Ion  Research  (HIC for FAIR) within the framework of the Landes-Offensive
zur Entwicklung Wissenschaftlich-Oekonomischer Exzellenz (LOEWE) program from the State of Hesse. J.M.T.-R. was supported by the U.S. Department of
Energy under Contract No. DE-FG-88ER40388. M.G. acknowledges support by the Deutsche Forschungs-Gemeinschaft (DFG) through the grant CRC-TR 211
``Strong-interaction matter under extreme conditions'' and from the ``Helmholtz Graduate School for Heavy Ion research''.
Computing services were provided by the Center for Scientific Computing (CSC) of the Goethe University Frankfurt.

\appendix

\section{1D+1 causal evolution of a charge fluctuation~\label{app}}

  In this Appendix we illustrate how a finite $\tau_Q$ helps to restore causality in the equations of the fluid~\cite{Kapusta:2014dja}. To do this we study the time evolution of an initial fluctuation of charge
  density in one spatial dimension. We consider the following two approximations for the induced electric current:
  
  \begin{enumerate}
   \item Noise-average of Eq.~(\ref{eq:ohm}) in the absence of electric fields and temperature gradients, i.e. 
   \be {\bf J}_Q = -\sigma_{\textrm{el}} \nabla \mu_Q \ . \ee
   \item Average of Eq.~(\ref{eq:ohmcausal}) with a finite $\tau_Q$ and the same conditions,
   \be \tau_Q \pa_t {\bf J}_Q + {\bf J}_Q = -\sigma_{\textrm{el}} \nabla \mu_Q \ . \ee
  \end{enumerate}
  
 In both cases we need to use the conservation of the electric charge $\pa_t n_Q = - \nabla \cdot {\bf J}_Q$, and relate $\mu_Q$ and $n_Q$ via the charge susceptibility $\chi_Q$,
  \be \label{eq:suscep} n_Q  = \left( \frac{ \pa n_Q}{\pa \mu_Q} \right)_T \mu_Q \equiv \chi_Q \mu_Q \ . \ee 

 Then, the first approximation gives the nonrelativistic evolution equation for an initial fluctuation $n_Q(0,{\bf x})$,
  \be \label{eq:diff} \pa_t n_Q(t,{\bf x}) - \sigma_{\textrm{el}} \chi_Q^{-1} \nabla^2 n_Q(t,{\bf x}) =0 \ . \ee
  This is the well-known diffusion equation. The second approximation, which contains a finite $\tau_Q$, gives the so-called Maxwell-Cattaneo equation
\be \label{eq:maxcat} \tau_Q \pa_t^2 n_Q(t,{\bf x}) + \pa_t n_Q(t,{\bf x}) - \sigma_{\textrm{el}} \chi_Q^{-1} \nabla^2 n_Q(t,{\bf x}) =0\ . \ee
    
  For a simple comparison of the two cases we take the 1+1 dimensional solution of these equations for an initial fluctuation in the form of $n_Q(0,x)=n_{Q,0} \delta(x)$. 
The analytical solutions of Eqs.~(\ref{eq:diff},\ref{eq:maxcat}) are known explicitly~\cite{Kelly}. They are supposedly to describe the time-space evolution of a fluctuation in the nonrelativistic, and 
the relativistic domains, respectively. 
  
  While we do not write the solution explicitly here (for more details and discussions we refer the reader to Ref.~\cite{Kelly}) we will plot them using the values of $\sigma_{\textrm{el}},\chi_Q$ and $\tau_Q$ obtained by SMASH in a particular case. To better illustrate the distinction between the relativistic and nonrelativistic cases we choose the values of these parameters for the single massless gas case at its highest temperature $T=0.3$ GeV
(cf. Fig.~\ref{fig:sigmamassless}).

\begin{figure}
\includegraphics[scale=0.4]{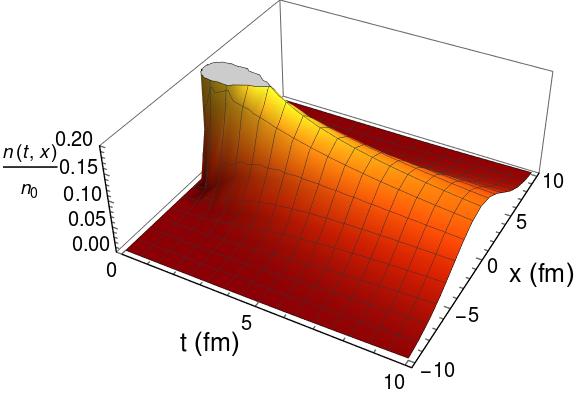}
\includegraphics[scale=0.4]{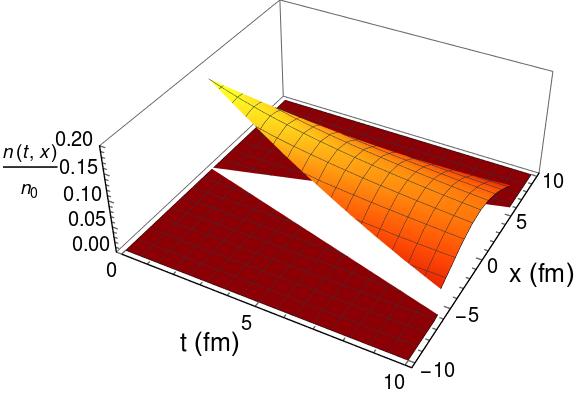}
\caption{Left panel: Solution in space and time of Eq.~(\ref{eq:diff}) for an initial perturbation $n_Q(0,x)=n_{Q,0} \delta(x)$. Right panel: 
Solution of Eq.~(\ref{eq:maxcat}) for the same initial condition.\label{fig:app}}
\end{figure}

The results are presented in Fig.~\ref{fig:app}, where we can observe that the solution using the Approximation 1 with $\tau_Q=0$ (left panel) has nonzero solution for all values of $x$, violating causality due to infinite propagation speeds. On the other hand, the Approximation 2 (right panel) which contains a finite relaxation time, gives a finite solution only within a compact support limited by a finite group velocity,
\be \label{eq:vel} v^2= \frac{\sigma_{\textrm{el}}}{\tau_Q \chi_Q} = \frac{V C(0)}{T \chi_Q} \ , \ee
where we have used Eq.~(\ref{eq:conducsimp}).

For this particular case, the group velocity corresponds to the speed of sound for massless gas~\cite{Kelly}. Substituting Eq.~(\ref{eq:C0massless}) into (\ref{eq:vel}), and using (\ref{eq:conducsimp}) and $n_Q=qen$ for the case of one species, we obtain
\be v^2 = \frac{ qe n_Q}{3T \left( \frac{\pa n_Q}{\pa \mu_Q} \right)_T} = \frac{1}{3} \ , \ee
where the equilibrium form of $n_Q$~has been used, cf. Eq.~(\ref{eq:C0massless}). The numerical value that we obtain using Eq.~(\ref{eq:vel}) reads $v^2 \simeq 0.331$, which can also be seen to be consistent with our solution in the right panel of Fig.~\ref{fig:app}.

\bibliography{ElecConducSMASH.bib}
%

\end{document}